\documentclass[useAMS,usenatbib]{mn2e}

\usepackage{txfonts}
\usepackage{graphicx}
\usepackage{times}
\usepackage{natbib}
\usepackage{subfigure}
\usepackage{footnote}
\usepackage{lscape}
\usepackage{multirow}
\usepackage{longtable}

\title[WR Population of NGC 5068]{The Wolf-Rayet population of the nearby
barred spiral galaxy NGC~5068 uncovered by VLT and Gemini}
\author[J.L.Bibby and P.A.Crowther]{J.L. Bibby$^{1,2}$\thanks{E-mail:
jbibby@amnh.org} and P.A. Crowther$^{2}$\\
$^{1}$Department of Astrophysics, American Museum of Natural History, 79th Street and Central Park
West, New York, NY 10024-5192 USA\\
$^{2}$Department of Physics \& Astronomy, University of Sheffield,
Hicks Building, Hounsfield Road, Sheffield, S3 7RH, UK}

\begin{document}

\date{}

\pagerange{\pageref{firstpage}--\pageref{lastpage}} \pubyear{2011}

\maketitle

\label{firstpage}

\begin{abstract} We present a narrow-band VLT/FORS1 imaging survey of
  the SAB(rs)cd spiral galaxy NGC~5068, located at a distance of 5.45
  Mpc, from which 160 candidate Wolf-Rayet sources have been
  identified, of which 59 cases possess statistically significant
  $\lambda$4686 excesses. Follow-up Gemini/GMOS spectroscopy of 64
  candidates, representing 40\% of the complete photometric catalogue,
  confirms Wolf-Rayet signatures in 30 instances, corresponding to a
  47\% success rate. 21 out of 22 statistically significant
  photometric sources are spectroscopically confirmed. Nebular
  emission detected in 30\% of the Wolf-Rayet candidates spectrally
  observed, which enable a re-assessment of the metallicity gradient
  in NGC 5068.  A central metallicity of log(O/H)+12 $\sim$ 8.74 is
  obtained, declining to 8.23 at $R_{\rm 25}$. We combine our
  spectroscopy with archival H$\alpha$ images of NGC 5068 to estimate
  a current star formation rate of
  0.63$^{+0.11}_{-0.13}$~$M_{\odot}$ yr$^{-1}$, and provide a
  catalogue of the 28 brightest HII regions from our own continuum
  subtracted H$\alpha$ images, of which $\sim$17 qualify as giant HII
  regions.  Spectroscopically, we identify 24 WC and 18 WN-type
  Wolf-Rayet stars within 30 sources since emission line fluxes
  indicate multiple Wolf-Rayet stars in several cases. We estimate an
  additional $\sim$66 Wolf-Rayet stars from the remaining photometric
  candidates, although sensitivity limits will lead to an incomplete
  census of visually faint WN stars, from which we estimate a global
  population of $\sim$170 Wolf-Rayet stars. Based on the
  H$\alpha$-derived O star population of NGC 5068 and
  N(WR)/N(O)$\sim$0.03, representative of the LMC, we would expect a
  larger Wolf-Rayet population of 270 stars.  Finally, we have
  compared the spatial distribution of spectroscopically confirmed WN
  and WC stars with SDSS-derived supernovae, and find both WN and WC
  stars to be most consistent with the parent population of Type Ib
  SNe.
\end{abstract}

\begin{keywords}
stars: Wolf-Rayet - Supernovae -  galaxies: stellar content, galaxies:
individual: NGC5068 - ISM: HII regions
\end{keywords}

\section{Introduction}

% replaced initial paragraph since no justification for "cluster" claim,
% chemical enrichment not only from massive stars
% observed in Lyman break galaxies, but there are local analogues (metallicities
% (close to lmc/smc).

Massive stars ($>$8\,M$_{\odot}$) dominate the radiative ionizing budget 
of star-forming galaxies, and contribute significantly to the
mechanical energy budget via their powerful stellar winds and ultimate 
death as core-collapse supernovae (ccSNe), plus the chemical enrichment 
of galaxies, especially for $\alpha$-elements. However, theoretically their 
evolutionary paths and precise fate remain unclear, arising from details 
of stellar winds, initial rotation and metallicity. Empirical results 
such as the ratio of blue and red supergiants, or the ratio of 
Wolf-Rayet (WR) to O stars provide sensitive tests of evolutionary models
which incorporate complex processes, such as rotation
\citep{LangerMaeder1995, Maeder2000}.

%Despite their extremely short lifetimes of only a few million years, these stars
%dominate the feedback to the local Interstellar Medium via their
%strong stellar winds and ultimate death as core-collapse supernovae
%(ccSNe) \citep{Crowther2007}

The conventional WR phase corresponds to the core-helium 
burning stage of massive stars ($\geq$20-25 M$_{\odot}$ in the Milky Way),
with a duration of only a few hundred thousand years \citep{Crowther2007}. 
WR stars possess winds densities which are an 
order of magnitude higher than O stars, producing a unique, broad 
emission-line spectrum. Spectroscopically, they are 
divided into WN and WC subtypes which are dominated by He\,{\sc ii}
$\lambda$4686 and C\,{\sc iii} $\lambda$4650 + C\,{ \sc iv}
$\lambda$5808  emission lines, respectively. Consequently,
WN and WC stars are  
associated with the products of the core hydrogen- (CNO cycle) and
helium-burning (3$\alpha$). Photometrically, WR stars 
cannot be distinguished from blue supergiants (BSG) via broad-band
imaging, although strong emission lines facilitate their detection 
via suitable narrow-band filters \citep{ms83}.

Follow-up spectroscopy of photometric candidates with 4m aperture
telescopes is routinely capable of determining the nature of
Wolf-Rayet stars in Local Group galaxies \citep{Neugent2011}. However,
beyond the Local Group 8m class telescopes are required. The
advantages of using telescopes such as the Very Large Telescope (VLT)
were first demonstrated by \citet{Schild2003} in their investigation
into the WR population of NGC~300.  Observations of massive stars
beyond $\sim$1 Mpc provide a broader range of galaxy morphological
types and metallicities than those available in the Local Group. In
addition, the much larger volume sampled enables progress in
empirically linking various flavours of core-collapse supernovae
(ccSNe) to progenitor stars.

%Over the past decade or so there has been a renewed interest in WR
%population surveys since WR stars are the leading candidate for the
%immediate progenitors of Type Ib (H-poor) and Type Ic (H+He-poor)
%core-collapse supernovae \cite{WoosleyBloom2006}. 

Massive stars with initial masses $\sim$8--20\,M$_{\odot}$ retain their
hydrogen envelope and end their lives in the red supergiant (RSG)
phase undergoing core--collapse and producing a H-rich Type II ccSN
\citep{Smartt2009}. This has been observationally confirmed
from archival broad-band pre-SN images e.g. SN 2003gd, which was
identified as an $\sim$8\,M$_{\odot}$ RSG \citep{Smartt2004}.
Stars with initial masses above $\sim$20\,M$_{\odot}$ are thought to
end their lives during the WR phase as 
H-poor Type Ib or H+He-poor Type Ic ccSNe
\citep{WoosleyBloom2006}. However, no direct detection of Type Ib/c SNe
progenitors have, to date, been established.

A greater understanding of the progenitors of ccSNe can be achieved
through observations of the environments in which the SNe
occur \citep{AndersonJames2008}. \citet{Kelly} found that H-rich Type II ccSNe follow the
distribution of the host galaxy light, whereas Type Ib and Ic ccSNe
are located in the brightest regions of the galaxy. Moreover, the
distributions of Type Ib and Ic ccSNe are different, suggesting that
they have different progenitors. If WN and WC stars are the
progenitors of Type Ib and Ic ccSNe respectively, then they should be
located in the same regions of the galaxy. \citet{Leloudas2010}
applied the same method to the spectroscopic WR surveys of M83 and NGC~1313
\citep{Hadfield2005, Hadfield2007}, confirming that the WN-Type Ib and
WC-Type Ic were the most likely progenitor scenarios, albeit with low
number statistics.

An alternative scenario for the production of Type Ib/c ccSNe has been
proposed by \citet{Podsialowski1992} in which intermediate mass stars
in binary systems lose their hydrogen and helium envelopes via Roche
Lobe overflow and/or common envelope evolution. This stripping of
outer layers would result in a low mass helium core which would have
WR-like emission, however, such low mass helium cores would be
completely hidden by their higher mass companions. This scenario
appears consistent with Type Ic SN 2002ap in M74, for which deep CFHT
broad-band images did not reveal any progenitor down to a limiting
magnitude of M$_{B}$\,=\,--4.2\,mag \citep{Crockett2007},
suggesting a binary system with a low mass C+O core of
$\sim$5\,M$_{\odot}$ \citep{Mazzali2002}.

At present, the relative statistics of core-collapse supernovae
\citep{Smarttetal2009, Li2011} favour both binary and single star
evolutionary channels \citep{Smith2011}. Our group are undertaking
photometric and spectroscopic surveys of Wolf-Rayet stars in a dozen
star-forming galaxies beyond the Local Group for the purpose of
studying massive stars in a variety of environments and to provide a
database of future Type Ib/c progenitors.

In this paper we investigate the WR population of NGC~5068, a face-on
barred spiral galaxy \citep[SAB(rs)cd]{deVau1991}, situated at a
distance of 5.45\,Mpc \citep{Herrmann2008} beyond the Centaurus A
group \citep{Karachentsev2007}. NGC~5068 is known to host WR stars,
since their spectroscopic signatures have been serendipitously
detected in observations of bright H\,{\sc ii} regions by
\citet{Rosa1986}. However, no systematic search for WR stars in
NGC~5068 has been undertaken and the stellar content of NGC 5068
remains largely unknown. Here, we present a comprehensive study of the
massive stellar population of NGC 5068, and compare the distribution
of WR stars relative to the giant H\,{\sc ii} regions in
NGC~5068. We constrain estimates of the star-formation rate
(SFR) of NGC~5068 which vary wildly, from
$\sim$0.35\,M$_{\odot}${\rm yr}$^{-1}$ to $\sim$3\,M$_{\odot}${\rm
  yr}$^{-1}$ \citep{Martin1997, Ryder1994}. Similarly, measurements of
the central metallicity of the galaxy range from LMC-like
metallicities \citep{Hodge1974} to super-solar \citep{Ryder1995}.

%Recent work by \citet{Modjaz2011} has extended the ccSNe distributions
%to include the metallicity of the supernova site, finding that Type Ib
%ccSNe are located in more metal-poor regions of the galaxy compared to
%Type Ic ccSNe. Moreover, the distribution of broad-lined (bl) Type Ic,
%often associated with long gamma-ray bursts (GRBs), are also found in
%more metal-poor regions which is consistent with previous work
%\citep{Modjaz2008}.

\begin{figure*}
%\centering
\subfigure[]{\includegraphics[width=1\columnwidth]{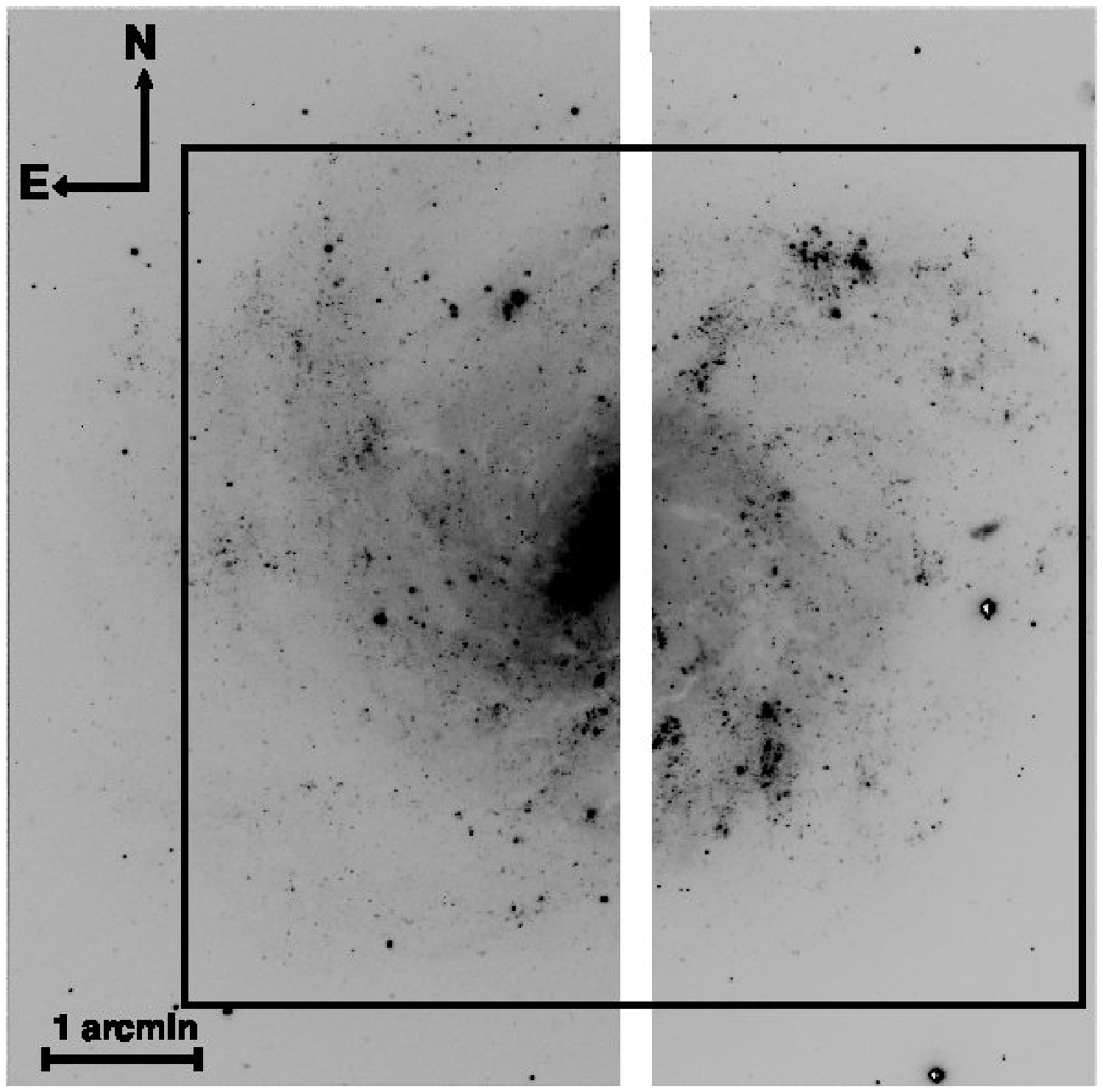}}
\subfigure[]{\includegraphics[width=1\columnwidth]{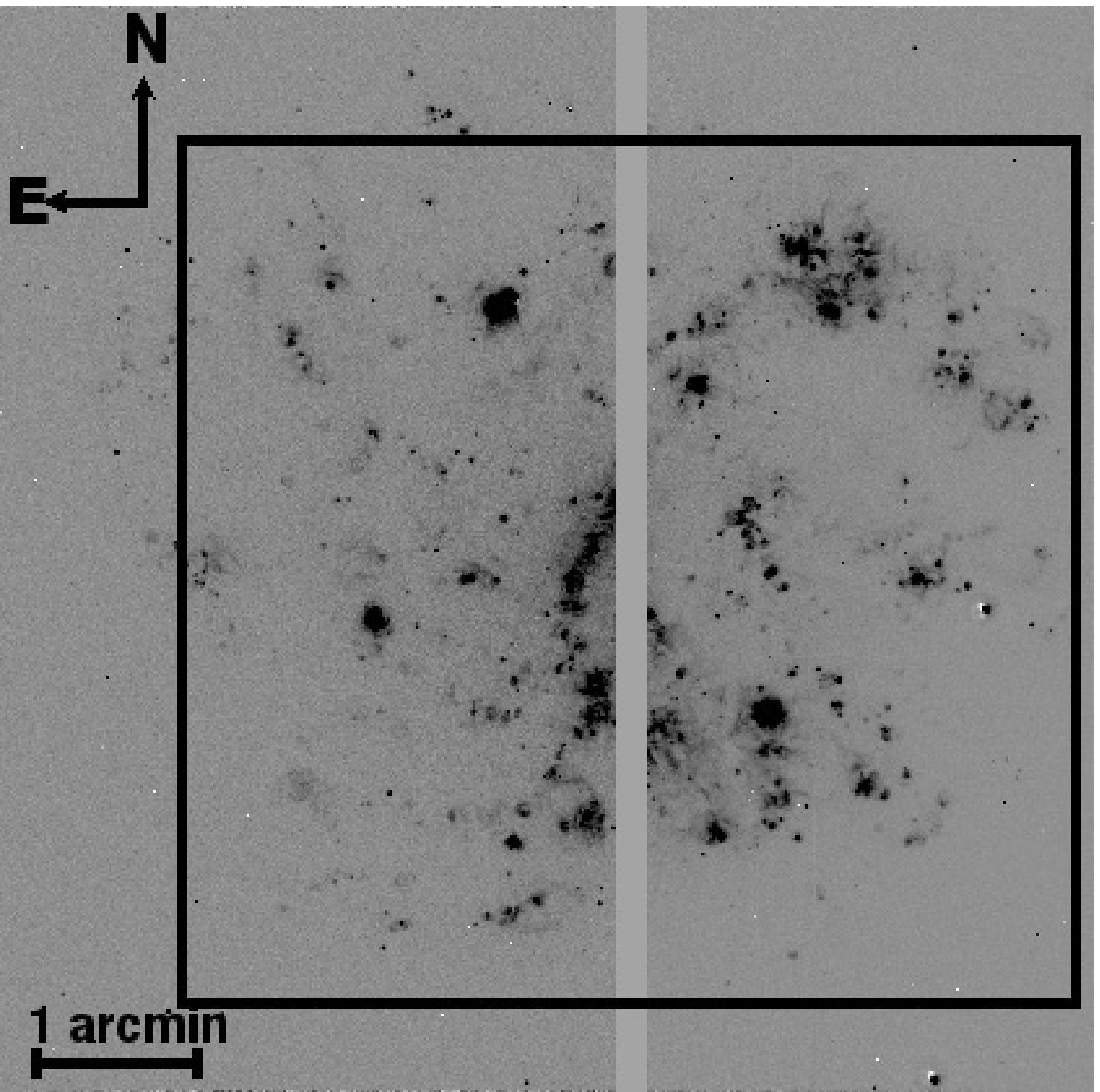}}
\caption{This figure shows a) He\,{\sc ii} image and b) the continuum
  subtracted H$\alpha$ image of NGC~5068 obtained using a single
  pointing of 6.8$\times$6.8 arcmin$^{2}$  with the VLT/FORS1 with the smaller 5.5
 $\times$ 5.5 arcmin Gemini/GMOS field of view indicated by the black
 square. The gap in the middle of the images represent the FORS1 chip
 gap hence we have no information about sources within this region.}
\label{fov}
\end{figure*}

This paper is organised as follows. In Section 2 we discuss the
observations, data reduction and source selection, while in Section 3
we analyse the nebular properties of H\,{\sc ii} regions in
NGC~5068. Section 4 determines the WR population and Section 5 the O
star population and SFR of NGC~5068. In Section 6 we assess the
global WR population based on the completeness of our
observations. Section 7 provides a comparison of results with both
NGC~7793 and predictions from evolutionary theory. Finally, Section 8
compares the spatial distribution of the WR stars in NGC~5068 with
those of ccSN; a summary follows in Section 9.

\section{Observations \& Data Reduction}

NGC 5068 has been imaged with the ESO Very Large Telescope (VLT) and
Focal Reduced/Low-dispersion Spectrograph \#1 (FORS1) covering a
field-of-view of 6.8 $\times$ 6.8 arcmin$^{2}$ with a plate scale of
0.25~arcsec~pixel$^{-1}$. Both broad- and narrow-band imaging were
obtained on 2008 April 7 under program ID 081.B-0289
(P.I. Crowther). In addition, the Gemini Multi-Object Spectrograph
(GMOS) on the Gemini-South telescope was used to obtain follow-up
spectroscopy in March-April 2009 under program ID GS-2009A-Q-20
(P.I. Crowther). The R150 grating was placed at a central wavelength
of 510nm and 530nm with a dispersion of
$\sim$3.5~$\AA$~pixel$^{-1}$. Further details of the observations can
be found in Table \ref{obs}.

\begin{table}
\begin{center}
\caption[Log of NGC5068 observations]{Observational log of
  VLT/FORS1 imaging and Gemini/GMOS spectroscopy of NGC~5068. Imaging
  was taken under program ID 081.B-0289 and spectroscopy under program
GS-2009A-Q-20.}
%\begin{tabular}{c@{\hspace{1mm}}c@{\hspace{1.5mm}}c@{\hspace{1mm}}c@{\hspace{1mm}}c@{\hspace{1mm}}}
\begin{tabular}{ccccc}
\\
\hline\hline
Date & Filter/  & $\lambda_{c}$ & Exposure & Seeing \\
   & mask ID  &     ($\AA$)        & time (s)  & (arcsec)    \\ 
\hline
\multicolumn{4}{c}{VLT/FORS1 Imaging}\\
\hline
2008-04-07   &    B high            & 4400   & 250             & 0.80    \\
         &    V high            & 5570     & 250             & 0.80    \\
         &    H$\alpha$         & 6563         & 250             & 0.85    \\
         &    H$\alpha$/4500    & 6665         & 250             & 0.75    \\
         &    He\,{\sc ii}      & 4684      & 3$\times$1025  & 0.80    \\
\vspace{2mm}
            &    He\,{\sc ii}/6500 & 4781       & 3$\times$1025  & 0.75    \\
\hline
\multicolumn{4}{c}{Gemini/GMOS Spectroscopy} \\
\hline
2009-04-03  &    MASK 1   &  5100/5300     & 6$\times$430    &  0.87 \\
 2009-03-31  &    MASK 2   &  5100/5300     & 6$\times$1400   &  0.87 \\
 2009-04-01  &    MASK 3   &  5100/5300     & 6$\times$1400   &  1.16 \\ 
\hline
\end{tabular}
\label{obs}
\end{center}
\end{table}

\subsection{Imaging \& Photometry}

FORS1 was used on 7 April 2008 to obtain narrow-band imaging of
NGC~5068 centered on $\lambda$4684 and $\lambda$4781 under good seeing
conditions of $\sim$0.8\arcsec. Three 1025~second
exposures were obtained for each filter while a single exposure of
250~seconds was used for additional narrow-band on-- and
off--H$\alpha$ images ($\lambda$6563 and $\lambda$6665,
respectively). Broad V- and B-band high-throughput images of
250~seconds each were also acquired on the same night to supplement
the narrow-band images. The field of view of the FORS1 images is
presented in Figure \ref{fov}, the black square indicates the region
covered by our spectroscopic observations which will be discussed
further in Section \ref{MOS}.

%\begin{figure}
%\centering
%\includegraphics[width=1\columnwidth]{}
%\caption{Continuum subtracted H$\alpha$ imaging of NGC~5068 obtained
%  with VLT/FORS, highlighting the H\,{\sc ii} regions .}
%label{ngc5068_ha}
%\end{figure}
Data reduction, including bias subtraction, flat fielding and image
combining were achieved using standard procedures in \textsc{iraf}
\citep{Tody1986}. Photometry was performed using the \textsc{daophot}
routine for the B, V, $\lambda$4684 and $\lambda$4781
images. Individual sources were fit with a gaussian point-spread
function (PSF) to determine their magnitude and its associated
error. 
%Sources were detected between m$_{4684}$\,=\,16 to 27\,mag,
%with most sources detected at m$_{4684}$\,=\,24.6\,mag as shown by the
%peak of the distribution in Figure \ref{detection_plot}a). 

\begin{figure}
%\centering
\includegraphics[width=0.73\columnwidth, angle=-90]{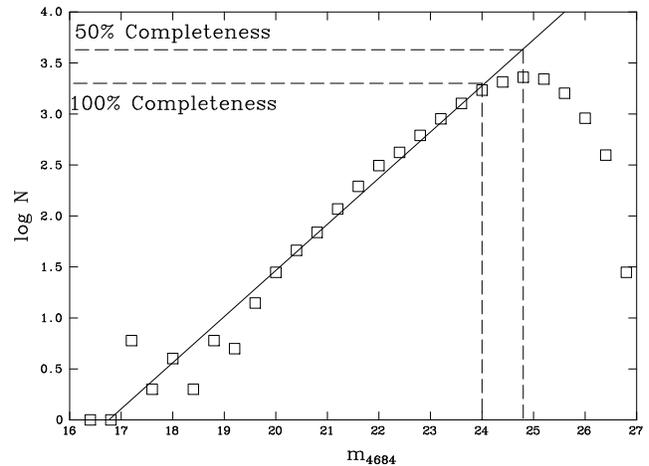}
\caption{Plot showing how the distribution of photometric
    sources in NGC~5068 varies with apparent $\lambda$4684 magnitude
    using 0.4\,magnitude bins. The 100\% and 50\% completeness limits
    are derived from this plot using the solid line which represents
    a linear fit to the brightest sources.}
\label{detection_comp}
\end{figure}

Figure \ref{detection_comp} shows the log distribution of the
  number of sources within each 0.4\,magnitude bin. Following
  \cite{bc10} we fit a power law to the bright end of the distribution
  using IDL, to assess the completeness of our imaging. The 100\%
  completeness limit is indicated by the turnover of the distribution
  where the observed data deviates from the extrapolated power law.
The 100\% completeness limit for our NGC~5068 narrow-band imaging
corresponds to m$_{4684}$\,=\,24.0\,mag. We define the 50\%
  completeness limit as the magnitude at which only 50\% of the
  sources predicted by the power law are detected, finding a 50\%
detection limit of m$_{4686}$\,=24.8\,mag for NGC~5068. The
significance of these detection limits is discussed further in
Section~\ref{detection_limits}. The detection limits for the
broad-band images are almost identical to those of the narrow-band
imaging.

%To derive the completeness limits of our imaging we extrapolated the
%logarithm of the distribution and applied an IDL fit. 

%\begin{figure}
%\centering
%\subfigure[]{
%\includegraphics[width=0.9\columnwidth, angle=-90]{completeness_5068.eps}
%}
%\subfigure[]{\includegraphics[width=0.8\columnwidth, angle=-90]{heII_50_comp.eps}}
%\caption{Apparent $\lambda$4684 magnitude distribution of photometric 
%sources in NGC~5068.}
%\label{detection_plot}
%\end{figure}

\begin{figure}
\centering
\includegraphics[width=0.9\columnwidth]{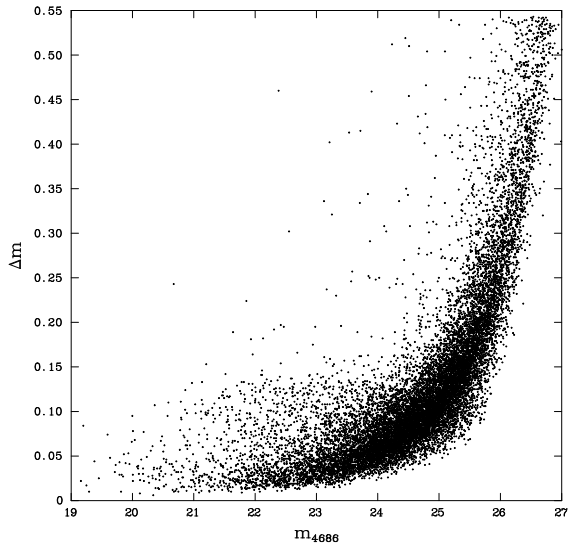}
\caption{Photometric errors as a function of apparent magnitude for all 
sources detected in the $\lambda$4684 VLT/FORS1 image of NGC~5068.}
\label{mag_error}
\end{figure}

Typical photometric errors for all images were approximately
$\pm$0.04\,mag for bright (m$_{\rm 4684}$\,=\,21\,mag) with
significantly higher errors of $\pm$0.20\,mag for the
faintest stars (m$_{\rm 4684}$\,=\,26\,mag) and $\pm$0.5\,mag
  for m$_{\rm 4684}$\,=\,27\,mag. Figure \ref{mag_error} shows the
distribution photometric errors for the $\lambda$4684 images. In some
instances, irrespective of the brightness of the source, severe
crowding makes a PSF fit inappropriate, so no photometry is derived
for some sources.

%The main distribution is due to intrinsic errors, 
%while the secondary distribution arises from
%crowding. For example, an isolated source of m$_{4684}$\,=\,25\,mag
%has typical errors of $\pm$0.15\,mag, however in a crowded region this
%increases to $\pm$0.4\,mag. 

Absolute calibration for the broad-band images was achieved from
observations of five stars within the standard star field SA~110-362
\citep{Stetson2000}. For the narrow-band images the spectrophotometric
standard star LTT~7987 \citep{Hamuy1994} was used to obtain
photometric zero--points since standard star fields are not
available. The associated error on the calculated zero--points is
$\pm$0.15\,mag.

\subsection{Source Selection}

The $\lambda$4684 narrow-band filter is coincident with the N\,{\sc
  iii}$\lambda$4640, He\,{\sc ii}$\lambda$4686 and C\,{\sc
  iv}$\lambda$4650 WR emission lines and the $\lambda$4781 filter is
placed on the continuum where there are no WR lines present. The
``blinking'' technique, pioneered by \citet{ms83}, is the easiest way of
identifying WR candidates and compares the $\lambda$4684,
$\lambda$4781 and continuum subtracted image ($\lambda$4684 minus the
$\lambda$4781 image) to reveal emission-line stars. 

In total 160 Wolf-Rayet candidates were identified in NGC~5068, which are
listed in Table \ref{sources}. Photometry was obtained for 95\% of the
candidates in at least the $\lambda$4684 filter. The few sources for which
$\lambda$4684 photometry was not obtained were either too faint,
extended or were located in a crowded region which makes reliable photometry
difficult; Very extended or crowded sources are noted in Table
\ref{sources}, as are three sources which lay at the very edge of the 
FORS1 CCD that naturally represent less secure candidates. 
Although all 160 candidates were identified from the `blinking' technique,
photometric errors indicates that only 59 sources are formally 3$\sigma$ detections.

To enable accurate mapping of the WR candidates an astrometric
solution was achieved by comparing the measured position of several
stars in the image to their positions found in the Guide Star
Catalogue. We found offsets of --0.6\,arcsec and +0.22\,arcsec for the
right ascension and declination, respectively, which have been applied
to the WR candidates listed in Table \ref{sources}.

% repeated text?

% We place an upper magnitude limit of
% m$_{\lambda4684}$\,=\,24.6\,mag, corresponding to our 100\%
% completeness limit, on the faint sources, however in practice these
% sources are likely to be somewhat fainter.

%\begin{landscape}

%\onecolumn
\begin{table*}
\centering
\caption{Catalogue of WR candidates in
  NGC~5608 ordered by Right Ascension.  Absolute magnitudes are
  derived using a distance of 5.45\,Mpc \citep{Herrmann2008}. Where
  spectra has been obtained the derived E(B-V) is used. For those
  sources with no nebular lines an average E(B-V)=0.259 mag is
  used. The spectral classification determined
  from spectroscopy is listed along with the sources association with
  a H\,{\sc ii} region. ``faint'' implies nebular--like H$\alpha$
  emission, ``bright'' implies a more point source H$\alpha$ emission
  while those in the brightest H\,{\sc ii} regions in NGC~5068 are
  referred to using ID numbers in Table \ref{ha}. The final column
  indicates what finding chart the source is located in in Appendix A.}
\label{sources}
\footnotesize
\begin{tabular}{
l@{\hspace{1mm}}
c@{\hspace{2mm}}
c@{\hspace{2mm}}
c@{\hspace{2mm}}
c@{\hspace{1mm}}
c@{\hspace{2mm}}
c@{\hspace{1mm}}
c@{\hspace{2mm}}
c@{\hspace{1mm}}
c@{\hspace{2mm}}
c@{\hspace{1mm}}
c@{\hspace{1mm}}
c@{\hspace{1mm}}
c}
\\
\hline\hline
ID & RA       &   Dec    & r/$R_{25}$ & m$_{V}$ & m$_B$--m$_{V}$ &
m$_{4684}$  & m$_{4684}$--m$_{4781}$  & E(B-V)  &  M$_V$ & M$_{4684}$
&  Spectral   &  HII region & Finding \\
     & J2000    & J2000    & & mag $\pm$ & mag $\pm$ & mag $\pm$ & mag
     $\pm$  & mag & mag & mag&  Type    & & Chart \\
\noalign{\smallskip}\hline
1 	        & 13:18:41.27	       & -20:59:53.1	& 1.20 & -- & --	               & 24.82 0.09	& --0.72 0.15  & 0.259  & -- & --4.79  & (WN)(1) & -- & 28 \\
2                & 13:18:41.27       & -21:04:36.9        & 1.07 & 21.68 0.07       & +1.76 0.07   & 22.82 0.02    & --0.16 0.10  & 0.259  & --7.80   & --6.79  &(no WR) & -- & 23 \\
3                &  13:18:43.15      & -21:01:38.4       & 0.79 & 22.06 0.29       & +0.27 0.29   & 22.29 0.02    & +0.01 0.09    & 0.259  & --7.42   & --7.32  & (no WR) & --& 24 \\
\smallskip
4             & 13:18:43.30       & -21:03:41.2        & 0.81 & 21.71 0.02       & +1.72 0.02   & 22.81 0.02    & --0.09 0.12  & 0.259  & --7.77    & --6.80  & (no WR) & faint & 23 \\
5             & 13:18:43.36       & -21:01:29.1        & 0.80 & \footnotesize{extended}  &      &                      & & 0.259  &               &              & (WR?)  & faint & 24\\
6 	        & 13:18:43.44	& -21:01:31.5	& 0.79 & 23.76 0.26	& +0.17 0.26  & 22.95 0.03	& --1.46 0.23 & 0.247  & --5.69   &  --6.62 & WN4--6(2) & faint & 24  \\ 
7 	        & 13:18:43.61	& -21:01:20.7	& 0.80 & 23.37 0.12	& +0.40 0.14  & 23.58 0.06	& --0.42 0.10 & 0.259  & --6.11   &  --6.03 &(no WR) & faint & 24  \\
\smallskip
8           & 13:18:43.64	& -21:01:24.6	& 0.79 & 23.19 0.07	& +0.06 0.07  & 23.30 0.03	& --0.03 0.07 &  0.259  & --6.29   &  --6.31 & no WR & faint & 24  \\
9           & 13:18:44.41       & -21:02:34.1        & 0.66 & 21.75 0.45        & --0.08 0.45 & 21.83 0.08    & --0.21 0.08 & 0.259   & --7.73   & --7.78  &(no WR)  & bright & 22  \\
10         & 13:18:44.75	& -21:01:04.3	& 0.78 & 24.30 0.09 & +0.64 0.11  & --	         	& --	        & 0.428   & --5.18   &  --       & WC4--5(1) & faint & 25  \\
11         & 13:18:44.85       & -21:00:50.4        & 0.82 & 21.68 0.02        & +0.17 0.06  & 22.09 0.07    & --0.09 0.09  & 0.259 & --7.80   & --7.52 &  (no WR) & HII \#3 & 25  \\
\smallskip 
12     & 13:18:45.09	& -21:01:03.8	& 0.76 & 22.81 0.06	& +0.16 0.07	& 22.35 0.02	& --0.95 0.25 & 0.284   & --6.75   &  --7.35  & WCE(1) & bright  & 25  \\
13     & 13:18:45.23	& -21:02:25.6	& 0.60 & 21.75 0.02	& +0.05 0.04	& 21.90 0.03	& --0.09 0.10 & 0.465   & --8.37   &  --8.45  & WCE(1)(?) & bright & 22   \\
14     & 13:18:45.56	& -21:02:33.5	& 0.58 & 21.57 0.22	& +0.12 0.22	& 21.72 0.02	& --0.22 0.26 & 0.259   & --7.91   &  --7.89  & no WR & HII \#4 & 22  \\
15   	& 13:18:45.77 	& -21:02:38.6	& 0.57 & --	        	& --		& 24.35 0.05	& --1.14 0.11 & 0.259   & --          &  --5.26  & (WN)(1) & --  & 22  \\
\smallskip  
16            & 13:18:45.84       & -21:02:30.5       & 0.56 & \footnotesize{extended}       &   &                     &                     & 0.259  &               &               & (WR?)             & HII \#4  & 22  \\
17         & 13:18:46.07	& -21:03:47.3       & 0.68 & 22.56 0.05	& +0.25 0.06	& 22.30 0.02	& +1.12 0.09   & 0.230  & --6.83   &  --7.21  & WC4-5(2)  & bright & 20  \\
18         & 13:18:46.12	& -21:03:25.0       & 0.62 & --       	        & 	--   	        & 24.59 0.05    & --0.55 0.09 & 0.259  & --          &  --5.02  & (WN)(1)   & --   & 20  \\
19           & 13:18:46.63	& -21:03:32.0       & 0.61 & 23.43 0.03	& +0.43 0.06	& 23.43 0.04    & --0.63 0.11 & 0.259  & --6.05   &  --6.18  & WC(1) & bright  & 20  \\
\smallskip
20        & 13:18:46.96       & -21:01:47.5        & 0.54 & 23.74 0.11        & --0.13 0.11  & 23.62 0.05   & --0.11 0.26 & 0.259  &                & --5.99   & (no WR)   & --  & 21  \\
21     & 13:18:47.01	& -21:00:33.8	& 0.77 & 21.72 0.21	& +0.89 0.21	 & 22.79 0.03   & --0.46 0.14 & 0.595  & --8.80    &  --8.03  & WNE(2)(?) & HII \#6 & 26  \\
22     & 13:18:47.09	& -21:00:25.3	& 0.80 & 23.15 0.45	& --0.01 0.45	 & --    	        & --                & 0.259  & --6.33     &  --         & no WR  & bright  & 27  \\
23     & 13:18:47.16	& -21:03:54.0	& 0.65 & 23.75 0.27	& --0.07 0.29	 & --                & --                & 0.259  & --5.73    &  --         & (WR?) & HII \#7  & 20 \\
\smallskip
24           & 13:18:47.24       & -21:00:36.5        & 0.75 & --                     &  --                & --                 & --               & 0.259  & --           &  --         &  (WR?) & --  & 26  \\
25     	& 13:18:47.438	& -21:00:19.7	& 0.81 & 20.29 0.10 	& +0.12 0.11	 & 20.44 0.04	 & --0.01 0.33 &0.259  &  --9.19   &  --9.17  & no WR & bright   & 27  \\
26           & 13:18:47.44       & -21:03:53.2        & 0.64 & 23.37 0.13        & --0.06 0.15  & 23.71 0.06    & --0.22 0.50 & 0.259 & --6.11    &  --5.90  & (no WR) & HII \#7   & 20  \\ 
27          & 13:18:47.47	& -21:01:57.4	& 0.48 & 23.39 0.09	& +1.83 0.15	 & 24.52 0.09	 & --0.38 0.09 & 0.259  &  --6.09  &  --5.09  & (no WR) & --  & 21  \\
\smallskip 
28             & 13:18:47.50        & -21:03:43.2        & 0.60 & 22.78 0.09        & +0.07 0.09   & 22.72 0.01    & --0.18 0.02 & 0.159  & --6.39    & --6.53   & (WN)(1) & faint  & 20  \\
29             & 13:18:47.50        & -21:03:28.3        & 0.55 & 25.19 0.05        & +0.01 0.12   & 24.80 0.11    & --0.30 0.31 & 0.259  &  --4.29   & --4.81   & (no WR) & --  & 20  \\
30 	       & 13:18:47.70	& -21:00:09.1	& 0.85 & 25.07 0.05	& -0.22 0.10	& 23.90 0.05	& --1.51 0.10  & 0.259  &  --4.41   &  --5.71 &  (WC)(1) & --  & 27  \\
31       & 13:18:47.74	& -21:02:14.6	& 0.44 & 21.95 0.03	& --0.02 0.08	& 22.03 0.09	& +0.04 0.24   & 0.259  & --7.53    &  --7.58  & no WR & --  & 21  \\	
\smallskip 
32    	& 13:18:47.78	& -21:02:19.2	& 0.44 & 23.76 0.05	& --0.22 0.06	& 22.48 0.02	& --1.27 0.45  & 0.259   & --5.72   &  --7.13   & (WN)(2) & --  & 21  \\
33    	& 13:18:48.05	& -21:03:52.3	& 0.61 & --		        & --		& --                &  --                &  --       & --          &  --          & (WR?) & --  & 20  \\
34 	& 13:18:48.06	& -21:00:48.7	& 0.66 & 18.61 0.04	& +0.67 0.07	& 19.59 0.07	& --0.08 0.48  & 0.259   & --10.87 &  -10.02  & no WR & HII \#9 & 26   \\
35 	& 13:18:48.12	& -21:00:37.7	& 0.70 & --	         	& 	--	        & --                 &                     & 0.259   & --           &  --         & WCE(2) & faint & 26 \\
\smallskip  
36          & 13:18:48.18	& -21:00:48.4	& 0.66 & --		        & --		& 21.54 0.10	& --1.02 0.10  & 0.259   & --          &  --8.07   & (WC)(1) & HII \#9  & 26  \\
37          & 13:18:48.18	& -21:00:38.9	& 0.70 & --                     &  --               & --                 & --                 & 0.259   & --          &  --          &  (WR?) & faint  & 26  \\
38          & 13:18:48.19        & -21:00:34.2        & 0.72 & --                    &  --               & --                 & --                 & 0.259   & --          &  --          &  (WR?) & faint  & 26  \\
39         & 13:18:48.29	& -21:00:33.4	& 0.72  & 22.59 0.06	& +0.21 0,08	& 22.75 0.05 	& --0.04 0.15   & 0.259  &  --6.89  & --6.86    & WN(1) & --  & 26  \\
\smallskip  
40           & 13:18:48.56        & -21:00:17.3        & 0.78 & 20.59 0.04       & +0.38 0.06    & 21.23 0.05   & --0.03 0.37   & 0.259  & --8.89    & --8.38   &(no WR) & HII \#10  & 27   \\
41        & 13:18:48.98	& -21:00:21.0	& 0.74 & 19.42 0.06	& +0.27 0.06	& 19.82 0.03	& --0.02 0.36   & 0.259  &  --10.06 & --9.79   & no WR  & HII \#11  & 27  \\
42        & 13:18:48.98	& -21:00:25.4	& 0.72 & 20.77 0.08	& --0.13 0.09	& 20.29 0.06	& --0.49 0.39   & 0.259  &  --8.71  & --9.32    &  (WC)(2) & faint  & 27  \\
43 	       & 13:18:49.15	& -21:02:33.1	& 0.35 & 22.90 0.05	& +0.09 0.05	& 22.42 0.02	& --0.65 0.06   & 0.259  &  --6.58  & --7.19  & WN5b(3) & --  & 15  \\
\smallskip
44     & 13:18:49.31	& -21:02:00.2	& 0.36 & 22.93 0.28	& --0.21 0.29   & 23.00 0.10	& --0.30 0.10   & 0.259  &  --6.55  & --6.66  & (WN)(1) & --  & 15  \\
45      & 13:18:49.36	& -21:00:21.6	& 0.73 & 19.53 0.16	& +0.38 0.16	 & 20.17 0.03	& 0.00 0.29	  & 0.259  &  --9.95  & --9.44  & no WR & HII \#12 & 27  \\
46        & 13:18:49.47        & -21:02:34.5        & 0.33 & 21.10 0.11        & +0.07 0.12    & 21.31 0.03   & +0.02 0.06    & 0.259  & --8.38   & --8.30  & (no WR)& bright  & 15  \\
47        & 13:18:49.49        & -21:03:55.9        & 0.57 & 21.56 0.21        & --0.10 0.22  & 21.55 0.09   & --0.38 0.15   & 0.259  & --7.92   & --8.06  & (no WR) & faint & 19 \\
\smallskip
48 	       & 13:18:49.60	& -21:03:38.3        & 0.49 & --                     & --                 & --                &    --               & 0.259  &  --         &  --        &  (WR?) & faint & 19  \\
49           & 13:18:49.69        & -21:03:19.9        & 0.42  &  20.05 0.07       & +0.31 0.07    & 20.33 0.01   & --0.03 0.07   & 0.259  & --9.43   & --9.28  & (no WR) & HII \#14 & 29  \\
50        & 13:18:49.72	& -21:03:41.7	& 0.50 & 21.46 0.07	& --0.13 0.10	 & 21.32 0.09	 & --0.27 0.09   & 0.259  &  --8.02  &  --8.29  & (WN)(2) & --  & 19  \\
51        & 13:18:49.78	& -21:04:06.8	& 0.61 & 22.64 0.08	& --0.34 0.09	 & 22.68 0.03	 & --0.59 0.49   & 0.259  &  --6.84  &  --6.93  & (WN)(1) & bright  & 19  \\
\smallskip 
52	       & 13:18:49.80	& -21:02:28.2	& 0.30 & 22.38 0.14	& --1.78 0.15	 & 20.99 0.02	 & --0.16 0.15   & 0.259  &  --7.10  &  --8.62  &(no WR) & HII \#13  & 15  \\
\hline
\end{tabular}
\end{table*}

\begin{table*}
{{\bf Table 2}(continued)} \\
\footnotesize
\begin{tabular}{
l@{\hspace{1mm}}
c@{\hspace{2mm}}
c@{\hspace{2mm}}
c@{\hspace{2mm}}
c@{\hspace{1mm}}
c@{\hspace{2mm}}
c@{\hspace{1mm}}
c@{\hspace{2mm}}
c@{\hspace{1mm}}
c@{\hspace{2mm}}
c@{\hspace{1mm}}
c@{\hspace{1mm}}
c@{\hspace{1mm}}
c}
\\
\noalign{\smallskip}\hline\hline
ID & RA       &   Dec    & r/$R_{25}$ & m$_{V}$ & m$_B$--m$_{V}$ &
m$_{4684}$  & m$_{4684}$--m$_{4781}$  & E(B-V)  &  M$_V$ & M$_{4684}$
&  Spectral & HII Region & Finding  \\
     & J2000    & J2000    & & mag $\pm$ & mag $\pm$ & mag $\pm$ & mag
     $\pm$  & mag & mag & mag&  Type    & & Chart \\
\noalign{\smallskip}\hline
53    & 13:18:49.80	& -21:02:26.5	& 0.30 & --     	        & 	--	        & 23.14 0.09	& --0.35 0.09   & 0.259  &  --         &  --6.47  & (WN)(1) & HII \#13  & 15 \\
54 	& 13:18:49.81	& -21:03:48.7	& 0.53 & 21.74 0.08	& +0.19 0.09	& 21.92 0.05	& --0.02 0.54   & 0.101  &  --7.25  &  --7.12  & no WR & --  & 19  \\
55 	& 13:18:49.84	& -21:03:20.7	& 0.42 & 22.11 0.05	& +0.08 0.09	& 22.43 0.05	& --0.23 0.15	 & 0.259  &  --7.37  &  --7.18  & no WR  & HII \#14 & 29\\
\smallskip 
56 & 13:18:49.90	& -21:02:30.3	& 0.30 & 23.28 0.20	& --0.12 0.21	& 22.32 0.02	& --2.00 0.37   & 0.259  &  --6.20  &  --7.29  & (WC)(1) & HII \#13  & 15 \\
57 & 13:18:49.98	& -21:03:56.8	& 0.56 & 23.64 0.05	& --0.59 0.07	& 23.47 0.06	& --0.72 0.06   & 0.259  &  --5.84  &  --6.14  & (WN)(1) & --  & 19  \\
58 & 13:18:49.99	& -21:03:23.7	& 0.42 & 19.42 0.14	& +0.05 0.15	& 19.19 0.08	& --0.39 0.49   & 0.259  &  --10.06 &  --10.42 & (no WR) & HII \#14  & 29  \\
59 & 13:18:50.03	& -21:02:17.5	& 0.29 & --	         	& --		& 22.14 0.01	& --1.62 0.35   & 0.259  &  --          &  --7.47  & WCpec?(?) & bright  & 15  \\
\smallskip
60  & 13:18:50.08	& -21:03:21.3	& 0.41 & 20.12 0.05	& +0.10 0.08	& 20.46 0.07	& --0.16 0.37   & 0.451  &  --9.96   &  --9.84  & no WR & HII \#14 & 29  \\
61 & 13:18:50.11	& -21:03:45.2	& 0.50 & 21.86 0.09	& --0.01 0.10	& 23.56 0.08	& --0.47 0.15   & 0.259  &  --7.62   &  --6.05  & (WN)(1) & --  & 19  \\
62 & 13:18:50.26	& -21:03:20.0	& 0.41 & 24.24 0.33	& --0.53 0.33	& 24.29 0.06	& --0.29 0.08   & 0.259  &  --5.24   &  --5.32  & (WN)(1) & bright  & 29\\
63  & 13:18:50.27	& -21:00:51.4	& 0.56 & 21.06 0.08	& +1.52 0.09	& 22.04 0.04	& --0.10 0.29	  & 0.259  &  --8.42   &  --7.57  & no WR & faint & 13 \\
\smallskip 
64 & 13:18:50.38	& -21:02:15.6	& 0.27 & 24.39 0.14	& --0.18 0.15	& 24.36 0.07	& --0.47 0.14   & 0.259  &  --5.09   &  --5.25   & (WN)(1) & faint  & 15\\
65 & 13:18:50.40	& -21:02:18.6	& 0.27 & 21.18 0.09	& +0.12 0.09	& 20.95 0.01	& --0.54 0.14   & 0.259  & --8.30    &  --8.66   & WC6(2) & bright  & 15  \\
66 & 13:18:50.41	& -21:03:42.9	& 0.48 & 24.56 0.12	& --0.42 0.14	& 23.95 0.06	& --0.41 0.14   & 0.259  & --4.92    &  --5.66   & (WN)(1) & --  & 19  \\
67 & 13:18:50.42    & -21:02:01.8  & 0.29 & 22.15 0.52        & --0.26 0.53  & 21.87 0.05   & --0.26 0.53   & 0.259  & --7.33    & --7.74    & (no WR) & bright & 15   \\
\smallskip 
68  & 13:18:50.50 & -21:02:02.2        & 0.28 & \multicolumn{3}{l}{\footnotesize{blended with A24}} &             & 0.259   &               &                & (WR?) & bright  & 15  \\
69  & 13:18:50.66	& -21:02:06.2	& 0.27 & --	    	        & --		& 22.04 0.03	& --0.49 0.03   & 0.259  & --           &  --7.57   & (WN)(2) & HII \#15   & 15 \\
70  & 13:18:50.67  & -21:02:07.6        & 0.26 & 20.37 0.13        & +0.02 0.13   & 20.32 0.03   & --0.10 0.11   & 0.259  &  --9.11   &  --9.29   & (no WR) & HII \#15  & 15  \\
71 & 13:18:50.80	& -21:01:08.6	& 0.46 & 20.70 0.21	& +1.70 0.21	& 21.79 0.04	& --0.09 0.09   & 0.259  & --8.78    &  --7.82   & no WR &  faint & 13 \\
\smallskip 
72    & 13:18:51.05    & -21:02:22.4          & 0.22 & --                     & --                & --                 & --                  & 0.259  & --           & --           & (WR?) & --  & 15  \\
73   & 13:18:51.16    & -21:04:09.3	        & 0.59 & 21.07 0.17	& +0.03 0.18	& 21.11 0.10	& --0.11 0.37	  & 0.277  &  --8.47   &  --8.57   & no WR  & HII \#16 & 18  \\
74    &  13:18:51.30   & -21:03:50.5          & 0.49 & --                     & --                & 23.04 0.05    & --0.17 0.07   & 0.259  &  --          & --6.57     & (no WR) & bright  & 18 \\
75     & 13:18:51.43     & -21:04:09.0	        & 0.58 & 21.78 0.10	& +0.08 0.12	& 21.79 0.05	& --0.25 0.11   & 0.259  &  --7.70   &  --7.82    &  (no WR) & HII \#16 & 18   \\
\smallskip 
76 & 13:18:51.50      & -21:04:08.3	        & 0.58 & 20.69 0.18	& +0.27 0.18	& 21.02 0.04	& --0.20 0.11   & 0.259  &  --8.79   &   --8.59   &  (no WR) & HII \#16 & 18\\
77 & 13:18:51.66      & -21:00:05.1	        & 0.74 & 20.41 0.16	& +1.36 0.17	& 21.32 0.05	& +0.03 0.08    & 0.259  &  --9.07   &   --8.29   & no WR  & faint  & 12  \\
78 & 13:18:51.68      & -21:01:18.4	        & 0.38 & 22.70 0.27	& --0.23 0.28	& 22.34 0.04	& --0.35 0.04	 & 0.259  &  --6.78   &   --7.27   & WN6(1) & HII \#17 & 14   \\
79 & 13:18:51.79      & -21:01:06.5	        & 0.44 & --      	        & 	--    	& 23.89 0.08	& --0.44 0.08   & 0.259  &  --         &  --5.72     &   (WN)(1) & --  & 13  \\
\smallskip 
80 & 13:18:51.78   & -21:03:37.3	& 0.42 & 25.25 0.06	& --0.33 0.10	& --                &     --              & 0.259  &  --4.23   &  --            & (WR?) & faint  & 17 \\
81 & 13:18:51.85	& -21:00:49.6	& 0.52 & 20.54 0.09	& +0.24 0.10	& 21.20 0.05	& +0.02 0.05	 & 0.259  &  --8.94   &  --8.41     & no WR   & HII \#18  & 13  \\
82  &13:18:51.86	& -21:01:16.1	& 0.39 & 19.74 0.18	& +0.05 0.18	& 19.63 0.05	& --0.25 0.07   & 0.259  &  --9.74   &  --9.98     &  (WN)(8) & HII \#17   & 14 \\
83 & 13:18:51.90	& -21:00:53.9	& 0.49 & 20.87 0.14	& +0.15 0.14	& 21.02 0.04	& +0.02 0.08    & 0.259  &  --8.61   &  --8.59     & no WR  & faint & 13  \\
\smallskip 
84	& 13:18:51.98 	& -21:03:41.2	& 0.44 & 24.52 0.02	& --0.21 0.04	& 22.72 0.02	& --2.35 0.23	 & 0.259  &  --4.96   & --6.89      & WC4--5(1)   & --  & 17  \\
85 	& 13:18:52.04 	& -21:01:14.9	& 0.39 & 21.80 0.06	& +0.35 0.07	& 22.23 0.03	& +0.06 0.39	 & 0.259  &  --7.68   & --7.38      & no WR  & HII \#17  & 14  \\
86 	& 13:18:52.15 	& -21:01:18.8	& 0.37 & 23.02 0.19	& --0.31 0.20	& 22.57 0.03	& --0.31 0.04   & 0.259  &  --6.46  &  --7.04      & (WN)(1) & bright  & 14  \\
87 	& 13:18:52.16 	& -21:01:14.2	& 0.39 & --                     &  --                &  --               & --                  & 0.259  &  --         &  --             & (WR?) & faint  & 14  \\
\smallskip 
88	& 13:18:52.19	& -21:01:21.7	& 0.35 & 24.41 0.12	& +0.15 0.14	& 23.03 0.03	& --2.80 0.06   & 0.259  &  --5.07   &  --6.58      & (WC)(1) & -- & 14  \\
89 	& 13:18:52.22	& -21:04:07.0	& 0.56 & --      	        & 	--    	& 22.42 0.03	& --2.54 0.13   & 0.259   &  --         & --7.19       & (WC)(1) & -- & 18 \\
90 	& 13:18:52.28	& -21:01:25.4  & 0.33 & 23.82 0.09	& --0.17 0.10	& 22.03 0.03	& --2.23 0.06	 & 0.259   &  --5.66   & --7.58       & WC4--5(2) & --  & 14 \\
91 	& 13:18:52.29	& -21:01:22.6	& 0.34 & 24.41 0.12	& +0.15 0.14	& 23.03 0.04	& --2.80 0.06   & 0.259   &  --5.07   & --6.58       & (WC)(1) & bright  & 14  \\
\smallskip 
92 	& 13:18:52.39	& -21:03:43.5	& 0.44 & 22.34 0.15	& +0.18 0.16	& 22.55 0.05	& --0.79 0.17   & 0.259   &  --7.14    & --7.06       & (WN)(2) & faint & 14 \\
93 & 13:18:52.57	& -21:00:57.8	& 0.46 & 21.42 0.19	& --0.14 0.02	& 21.33 0.06	& --0.11 0.35	 & 0.170    &  --7.79    & --7.96      & WNE(1) & bright  & 13  \\
94 	& 13:18:52.71	& -21:03:04.3	& 0.25 & 25.14 0.08	& --0.71 0.09	& 22.55 0.02	& --3.36 0.05	 & 0.259    &  --4.34    & --7.06      & WC7(1) & --  & 16  \\
95 	& 13:18:52.72	& -21:03:27.6	& 0.36 & -- & 	--     	& 21.99 0.09	& +0.12 0.28	 & 0.259    &  -- & --7.62      & no WR & faint  & 17 \\
\smallskip
96 	& 13:18:52.95	& -21:02:54.3	& 0.19 & 23.46 0.31	& --0.15 0.32	& 22.39 0.02	& --1.20 0.19	& 0.259     &  --6.02    & --7.22      & (WN)(2) &  bright  & 16 \\
97    & 13:18:53.06  & -21:03:56.8        & 0.50 & 21.44 0.29         &+1.98 0.29    & 22.64 0.04   & --0.12 0.09  & 0.259     &  --8.04    & --6.97      &  (no WR) & bright  & 18 \\
98 	& 13:18:53.06	& -21:03:25.7	& 0.34 & --             	& 	--    	&                     & --                 & 0.259     &  --           & --             & (WR?) & bright  & 17  \\
99 	& 13:18:53.08	& -21:03:22.7	& 0.33 & 23.66 0.04	& +0.53 0,05	& --		& --		& 0.259     &  --5.82    & --             & WN6(3) & bright  & 17  \\
\smallskip
100 	& 13:18:53.11	& -21:03:30.9	& 0.37 & 23.17 0.33	& +0.12 0.33	& 23.11 0.05	& --0.21 0.07   & 0.259   &  --6.31    & --6.50       & (WN)(1) & bright  & 17  \\
101 & 13:18:53.13	& -21:03:36.6	& 0.40 & 19.69 0.05	& +0.21 0.06	& 19.94 0.04    & +0.07 0.46	 & 0.259   &  --9.79     & --9.67      & no WR & faint & 17  \\
102  & 13:18:53.14  & -21:00:32.2  & 0.57 & 22.03 0.22        & +0.23 0.23   & 22.38 0.06    & --0.13 0.39  & 0.259   & --7.45      & --7.23      & (no WR) & bright & 10   \\
103 & 13:18:53.23	& -21:02:43.5	& 0.14 & 22.39 0.09	& --0.17 0.11	& 22.36 0.02	& --0.18 0.45   & 0.027   &  --6.37    & --6.42      & no WR & HII \#19   & 16  \\
\smallskip
104 & 13:18:54.15	& -21:02:03.8 	& 0.09 & edge of chip	& 	--	         & --		&    --               & 0.147   &  --           & --             & WN(1) & --  & 8  \\
105 & 13:18:54.15	& -21:03:26.7       & 0.34 & edge of chip 	& 	--	         & 	--            & 	--             & --        &  --           & --             & no WR  & bright  & 7  \\
106 & 13:18:54.18	& -21:00:27.9 	& 0.59 & 23.21 0.12	& --0.19 0.13	 & 22.80 0.04	& --0.57 0.51	 & 0.087   &  --5.74    & --6.19       & WN7(1) & bright  & 10  \\
107 	& 13:18:54.19	& -21:03:07.6 	& 0.24 & edge of chip      &      --              &    --           &        --           & 0.259   &  --           & --              & (WR?) & bright  & 7 \\
\smallskip
108 & 13:18:54.22	& -21:00:32.1	& 0.56 & 20.88 0.05        & +0.01 0.05  	 & 20.99 0.03	& --0.05 0.18   & 0.259   &  --8.60    & --8.62 & no WR & faint  & 10 \\
\hline
\end{tabular}
\end{table*}

\begin{table*}
{{\bf Table 2}(continued)} \\
\footnotesize
\begin{tabular}{
l@{\hspace{1mm}}
c@{\hspace{2mm}}
c@{\hspace{2mm}}
c@{\hspace{2mm}}
c@{\hspace{1mm}}
c@{\hspace{2mm}}
c@{\hspace{1mm}}
c@{\hspace{2mm}}
c@{\hspace{1mm}}
c@{\hspace{2mm}}
c@{\hspace{1mm}}
c@{\hspace{1mm}}
c@{\hspace{1mm}}
c}
\\
\noalign{\smallskip}\hline\hline
ID & RA       &   Dec    & r/$R_{25}$ & m$_{V}$ & m$_B$--m$_{V}$ &
m$_{4684}$  & m$_{4684}$--m$_{4781}$  & E(B-V)  &  M$_V$ & M$_{4684}$
&  Spectral  & HII Region & Finding \\
     & J2000    & J2000    & & mag $\pm$ & mag $\pm$ & mag $\pm$ & mag
     $\pm$  & mag & mag & mag&  Type  & & Chart   \\
\noalign{\smallskip}\hline
109   & 13:18:54.24        & -21:02:04.6       & 0.08  & 20.65 0.26        & +0.57 0.27   & 20.85 0.10   & --0.25 0.16  & 0.259   & --8.83     & -- 8.7  & (no WR) & bright & 8 \\ 
110 	& 13:18:54.25 	& -21:03:06.5	& 0.24 & 24.40 0.03	& --0.28 0.10	& 21.15 0.02	& --0.15 0.03	& 0.102   &  --5.08 & --8.46 & (no WR) & faint  & 7  \\
111 	& 13:18:54.27 	& -21:02:55.1	& 0.18 & 25.04 0.12	& --0.57 0.14	& 23.42 0.03	& --1.50 0.06	& 0.259   &  --4.44    & --6.19 & WNL(1) & --  & 7 \\
\smallskip 
112        & 13:18:54.31	        & -21:03:08.2 	& 0.25 &  --                    & --                 & 22.14 0.04	& --                  & 0.102   &  --    & --6.91 & WC(1) & faint  & 7 \\
113    	& 13:18:54.38 	& -21:00:29.3	& 0.58 & --              	& 	--	        & 	--            & --                  & 0.259   &  --           & --        & (WR?) & faint  & 10  \\
114          & 13:18:54.49       & -21:02:13.3       &  0.04 & \footnotesize{crowded} &         &                      &                      &            &                   &            &  (WR?)  & bright   & 8  \\
115   	& 13:18:54.60 	& -21:02:07.9	& 0.07 & 19.87 0.05	& +0.15 0.08	& 19.99 0.09	& --0.03 0.21   & 0.259  & --9.61      &  --9.62   & no WR & faint  & 8  \\
\smallskip 
116       & 13:18:54.67	& -21:03:10.8	& 0.26 & 18.91 0.13	& +0.05 0.14	& 18.99 0.06	& --0.16 0.44   & 0.259  & --10.57    &  --10.62  & no WR  & HII \#20 & 7  \\
117       & 13:18:54.67	& -21:03:06.1	& 0.24 & 21.20 0.22	& +0.20 0.22	& 21.37 0.02	& --0.15 0.12   & 0.259  & --8.28      &  --8.30    & B0-B3 & bright  & 7  \\
118        & 13:18:54.77        & -21:04:08.8        & 0.56 & 23.67 0.18        & --0.02 0.18  & 22.68 0.03   & --1.34 0.11  & 0.259   & --5.81      &  --6.93  &  (WN)(2) & faint  & 5 \\
119      & 13:18:54.82	& -21:03:12.8	& 0.27 & 22.68 0.23	& --0.39 0.23	& 21.62 0.02	& --1.34 0.30   & 0.259  & --6.80      &  --7.99    & (WC)(1) & bright  & 7  \\
\smallskip 
120            & 13:18:54.83       & -21:0.3:26.0       & 0.34  & 21.88 0.09        & +0.05 0.09    & 21.99 0.03   & +0.03 0.08    & 0.259   & --7.60      & --7.62     & (no WR) & --   & 7  \\ 
121           & 13:18:54.90       & -21:04:02.8        & 0.53 & \footnotesize{extended} &          &                     &                      &             &                   &                & (WR?) & HII \#21 & 5  \\
122            & 13:18:54.96       & -21:0.4:55.3       & 0.81 & 21.30 0.09        & +1.60 0.10   & 22.44 0.03    & --0.02 0.34   & 0.259   &  --8.18     & --7.17     & (no WR) & --  & 11  \\
123      & 13:18:54.97       & -21:04:06.2	       & 0.55 & 21.95 0.04	       & +0.03 0.05   & 22.02 0.04	& --0.08 0.13	 & 0.259   & --7.53      &  --7.59    & no WR & bright  & 5\\
\smallskip 
124 	        & 13:18:55.11      & -21:04:05.0	       & 0.55 & 23.43 0.22	       & -0.22 0.23	& 22.06 0.01	& --1.54 0.10   & 0.259  & --6.06       &  --7.55  & (WC)(1) & faint  & 5  \\
125	& 13:18:55.15      & -21:02:57.1	       & 0.20 & 22.18 0.08        & +0.06 0.08	& 21.93 0.03	& --0.43 0.04	 & 0.259   & --7.30       &  --7.68  & WC(1) & bright & 7   \\
126 	        & 13:18:55.35      & -21:02:32.3        & 0.09 & 21.81 0.11	       & +0.42 0.13	& 21.83 0.09	& --0.32 0.09   & 0.259  & --7.67        &  --7.78  & (WN)(2) & bright  & 8 \\
127 	        & 13:18:55.42      & -21:04:13.1	       & 0.59 & --                     & 	--	        & 24.08 0.05	& --1.16 0.14   & 0.259  & --               &  --5.53  &  (WN)(1) & --  & 5  \\
\smallskip
128   & 13:18:55.53	& -21:04:06.7	& 0.56 & 24.04 0.06	& +0.26 0.08	& extended      &         --          & 0.259  & --5.44        &  --         & WC4--5(1) & bright  & 5  \\
129   & 13:18:55.66	& -21:03:39.6	& 0.42 & --	   	        & --		& 22.83 0.03	& --0.55 0.32   & 0.358  & --               &  --7.14   & no WR & bright  & 6  \\
130 	& 13:18:55.79	        & -21:03:46.6	& 0.46 & 22.21 0.05	& +0.55 0.08	& 22.84 0.09	& --0.46 0.33   & 0.259  & --7.27        &  --6.77   & no WR & bright  & 6  \\
131 	& 13:18 56.02  	& -21:02:44.8	& 0.17 & 23.46 0.31	& --0.15 0.32	& 22.39 0.02	& --1.20 0.19	 & 0.180   &  --5.78       & --6.94    & WC4(1) & bright  & 2  \\
\smallskip
132  & 13:18:56.32	& -21:04:36.3	& 0.73 & 21.72 0.18	& --0.06 0.19	& 22.01 0.06	& --0.54 0.06   & 0.259   &  --7.76      & --7.60    & (no WR) & bright  & 4  \\
133 	& 13:18 56.52	& -21:02:04.7	& 0.15 & 23.87 0.09	& +0.18 0.10	& 22.89 0.04	& --1.55 0.08	 & 0.259   &  --5.61       & --6.72    & WC4(1) & faint & 2  \\
134   & 13:18:56.77 & -21:04:13.4	& 0.62 & 21.13 0.15	& +0.58 0.16	& 21.12 0.02	& --1.24 0.06	 & 0.181   &  --8.11       & --8.21    & WC4(3) &HII \#24 & 4  \\
135 	& 13:18:56.82	& -21:02:13.7	& 0.16 & 20.12 0.09        & +1.49 0.1	& 21.04 0.04	& --0.05 0.15	 & 0.259	 &  --9.36      & --8.57    & no WR & faint  & 2  \\
\smallskip
136 	       & 13:18:56.82	& -21:04:41.6	& 0.76 & 22.74 0.09	& +0.34 0.12	& 22.65 0.06	& --1.01 0.06	& 0.259    &  --6.74       & --6.96    & no WR & faint  & 4\\
137           & 13:18:56.91        & -21:03:10.7        & 0.32 & 21.97 0.13        & --0.02 0.13  & 21.99 0.02   & --0.05 0.09  & 0.259   & --7.51        & --7.62    & (no WR) & bright  & 6  \\
138           & 13:18:56.98        & -21:04:13.1        & 0.62 & 20.38 0.41        & +0.12 0.42   & 20.46 0.03   & --0.27 0.45  & 0.259   &  --9.10        & --9.15    & (no WR) & HII \#24  & 4  \\
139        & 13:18:57.09	       & -21:00:46.1	& 0.51 & --                     & --                 & --                & --                 & 0.259   &  --              & --            & (WR?) & HII \#25 & 9  \\
\smallskip 
140             & 13:18:57.17	& -21:03:25.3	& 0.39 & 23.37 0.21 	& +0.08 0.21	& 23.31 0.04	& --0.25 0.06   & 0.259   &  --6.11      & --6.30     & (WN)(1) & faint &  6\\
141      & 13:18:57.20	& -21:02:07.8	& 0.19 & 21.70 0.06        & --0.11 0.07	& 21.50 0,03	& --0.28 0.06	 & 0.259   &  --7.78       & --7.91     & no WR & bright  & 2  \\
142             & 13:18:57.20	& -21:00:47.8	& 0.50 & 21.25 0.15	& --0.15 0.15	& 20.88 0.06	& --0.29 0.13   & 0.259   &  --8.23      & --8.73     & (no WR) & HII \#25  & 9  \\
143        & 13:18:57.63        & -21:03:22.0        & 0.40 & 22.34 0.03        & +0.14 0.09   & 22.25 0.10   & --0.56 0.43   &  0.259  & --7.14        & --7.36    &  (no WR) & faint  & 6  \\
\smallskip 
144    & 13:18:57.65	& -21:04:41.9	& 0.78 & 23.78 0.04	& +0.24 0.07	& 22.28 0.02	& --2.58 0.41   & 0.259   &  --5.70       & --7.33    & (WC)(1) & bright   & 4  \\
145     & 13:18:57.98	& -21:02:28.4	& 0.24 & 23.06 0.05	& --0.07 0,06	& 22.08 0.02	& --1.11 0.17	 & 0.259   &  --6.42       & --7.53    & WC6(1) & HII \#26 & 2  \\
146     & 13:18:58.04	& -21:00:58.4	& 0.47 & 19.94 0.06	& +1.58 0.07	& 20.98 0.03	& --0.09 0.06   & 0.259   &  --9.54       & --8.63    & no WR & bright  & 9  \\
147       & 13:18:59.01	& -21:04:08.6	& 0.66 & 24.00 0.06	& +0.35 0.08	& 23.43 0.03	& --1.06 0.13   &  0.259  & --5.48        & --6.18    & (WN)(1) & -- & 3   \\
\smallskip 
148      & 13:18:59.58        & -21:02:09.4         & 0.34 & 23.87 0.09        & +0.18 0.10   & 23.52 0.07   & --0.68 0.32    & 0.259   & --5.61       & --6.09    &  (WN)(1) & --  & 2  \\
149      & 13:19:00.02	       & -21:04:44.4	& 0.86 & 22.62 0.40	& +0.67 0.40	& 23.23 0.03	& --0.53 0.30	  &  0.125  & --6.45        & --5.90   & WNE(1) & bright  & 3  \\
150      & 13:19:00.30	       & -21:04:46.2	& 0.88 & 24.33 0.17	& +0.14 0.18	& 22.72 0.02	& --2.72 0.08	  &  0.259  & --5.15        & --6.89   & WC4--5(1) & faint  & 3  \\
151    & 13:19:00.77	       & -21:02:25.2	& 0.42 & 25.04 0.12	& --0.57 0.14	& 23.42 0.03	& --1.50 0.06   &  0.259   & --4.44       & --6.19   & WNE(1) & --  & 2  \\
\smallskip
152      & 13:19:00.79	       & -21:02:47.3	& 0.45 & 19.58 0.12	& +0.40 0.14	& 20.79 0.09	& +0.37 0.09    &  0.259  & --9.90         & --8.82 & (no WR) & HII \#28  & 2  \\
153      & 13:19:01.98	       & -21:00:36.7	& 0.70 & 20.89 0.11	& --0.14 0.12	& 20.76 0.06	& --0.28 0.21   &  0.259  & --8.59         & --8.85 & (no WR) & bright  & 1 \\
154      & 13:19:02.02	       & -21:02:44.2	& 0.52 & 24.98 0.12	& --0.40 0.15	& 24.53 0.10	& --0.72 0.11   & 0.259   & --4.50         & --5.08 & (WN)(1) & --  & 2  \\
155       & 13:19:02.19	       & -21:01:15.3	& 0.59 & 19.82 0.21	& +1.49 0.21	& 20.80 0.04	& --0.05 0.53   & 0.259   & --9.66         & --9.79 & no WR & bright & 1    \\
\smallskip  
156      & 13:19:02.24         & -21:01:11.5	& 0.60 & 21.59 0.08	& +0.24 0.08	& 21.81 0.04	& +0.09 0.08	 & 0.259   & --7.89          & --7.80 & no WR & --  & 1  \\
157        & 13:19:02.74          & -21:01:09.8        & 0.63 & 22.48 0.49        & --0.33 0.49  & 22.21 0,05   & --0.32 0.31  & 0.259   & --7.00          & --7.40  & (no WR) & bright & 1   \\
158         & 13:19:02.93          & -21:03:52.0        & 0.76 & 22.90 0.16        & --0.05 0.16  & 22.75 0.02   & --0.17 0.12  & 0.259   & --6.58          & --6.86  & (no WR) & faint   & 3  \\
159        & 13:19:05.09          & -21:01:27.8        & 0.73 & 25.45 0.12        & --0.47 0.16  & 24.17 0.08   & --1.21 0.15  & 0.259   & --4.03          & --5.44  & (WN)(1) & --  & 1  \\
\smallskip 
160         & 13:19:05.13       & -21:02:26.4        &  0.70 & 23.54 0.02        & +0.05 0.08    & 23.53 0.06   & --0.31 0.33  & 0.259   & --5.94           & --6.07  & (no WR) & faint & 2   \\
\hline
H1 	& 13:18:58.05	& -21:01:45.9	& 0.28 & 23.24 0.35 & +0.01 0.35	& 23.48 0.05	& +0.03 0.30    & 0.258   &  --6.23 & --6.13  & HII region & bright \\
H2 	& 13:18:49.50	& -21:01:52.6	& 0.37 & 23.38 0.08 & --0.07 0.09	& 23.63 0.03	& +0.04 0.07    & 0.258   &  --6.10 &  --5.98 & HII region & bright\\
H3 	& 13:18:57.01	& -21:04:25.9	& 0.69 & 23.25 0.08	& --0.11 0.09	& 23.29 0.06	& --0.02 0.10	& 0.181    &  --5.99   & --5.95  & HII region & bright\\
H4 	& 13:18:55.75	& -21:01:37.0	& 0.23 & 21.80 0.08 & +0.31 0.08	& 22.09 0.03	& +0.05 0.14	& 0.259    & --7.68 &  --7.52  & HII region & bright\\
\end{tabular}
\end{table*}

%\end{landscape}

%\twocolumn

\subsection{MOS Spectroscopy}\label{MOS}

Multi-Object Spectroscopy (MOS) of WR candidates in NGC~5068 was
obtained in March-April 2009 using the Gemini Multi-Object
Spectrograph (GMOS) at the Gemini-South telescope in
Cerro Pachon, Chile.

Since the $\lambda$4684 and $\lambda$4781 narrow-band imaging were
obtained from the VLT\footnotemark\footnotetext{$\lambda$4684 and
  adjacent continuum filters were subsequently acquired for
  Gemini/GMOS}, broad-band (g' filter) pre-imaging was obtained using
GMOS to aid the MOS mask design. The field-of-view of GMOS is
5.5$\times$5.5 arcmin compared to the 6.8$\times$6.8 arcmin VLT field
(Figure \ref{fov}) hence some of the WR candidates lay beyond the GMOS
field. The GMOS field of view was chosen to maximise the number of WR
candidates and is overlaid on the VLT field in Figure \ref{fov}; only
six sources were outside the GMOS field.

Three MOS masks were designed using 0.75 arcsec slits and the R150
grating with a dispersion of $\sim$3.5\,$\AA$\,pixel$^{-1}$. The
spectral coverage was typically from $\sim$3900--9000\,$\AA$ to
include all of the WR and nebular diagnostic lines. Data long-ward of
$\sim$7000\,$\AA$~ suffered from fringing effects and second order
contamination so was unreliable, however this is not a concern as no
diagnostic lines lie within this region.

For the first MOS mask (\#1) six 430~s exposures were obtained under
good seeing conditions which over the duration of the observations was
an average of 0.87\arcsec. WR candidates included in MOS
mask \#1 had the brightest $\lambda$4684--$\lambda$4781
photometric excesses requiring a S/N$\sim$3--5 for detection. MOS
masks \#2 and \#3 had six longer exposures of 1400~s each, producing a
higher signal-to-noise of S/N$\sim$10 to detect the fainter WR
candidates. Seeing conditions for MOS mask \#2 were similar to the
mask \#1, although conditions deteriorated to
1.16~\arcsec~ for mask \#3.

64 out of the 160 WR candidates were spectroscopically observed, 
($\sim$40\%), a similar percentage to the 3$\sigma$ 
photometrically significant sources (23 from 59 cases).
To fill the gaps in the MOS mask design, four H\,{\sc ii}
regions, identified from our H$\alpha$ images were added, allowing
nebular properties to be investigated. Additional targets were
spectroscopically observed in regions of NGC~5068 where we were unable
to identify suitable WR candidates due to the
FORS chip gaps, two of which were spectroscopically observed.

Spectroscopic data were reduced and extracted using standard
procedures within the Gemini \textsc{iraf} package. Spectra were
wavelength calibrated using observations of an internal CuAr arc lamp
with each of the three MOS masks. Absolute flux calibrations were
achieved from observations of the spectrophotometric standard star LTT
7987 \citep{Hamuy1994} using the \textsc{starlink} package
\textsc{figaro}.

\begin{figure}
\includegraphics[width=0.73\columnwidth, angle=-90]{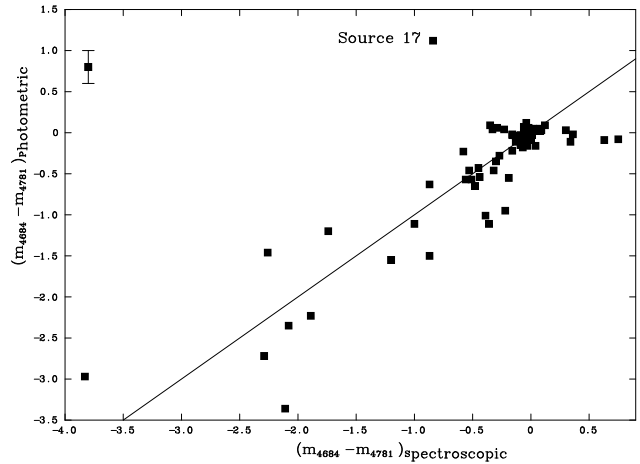}
\caption{Comparison of spectroscopic and photometric $\lambda$4684 excesses of NGC~5068
sources. The solid line indicated where the excesses
are in agreement. The error bar ($\pm$0.2\,mag) shown is typical of
faint sources ($\sim$24-26\,mag), however errors are lower for brighter sources.}
\label{outliers}
\end{figure}

In order to assess the reliability of our photometry we convolved the
observed flux calibrated object spectra with the response function of
the interference filter used in the narrow-band imaging to determine a
spectroscopic magnitude for each source; a comparison of which is
shown in Figure~\ref{outliers}. Most sources are in reasonable
agreement, with the exception of source \#17, the
m$_{\lambda4684}$--m$_{\lambda4781}$ excess of $+$1.12 is inconsistent
with the WC subtype determined from spectroscopy (see Section
\ref{spec_wc}). This source was included as a candidate following
visual inspection of the continuum subtracted image which clearly
showed the presence of WR emission, hence photometry for this source
is unreliable.

\section{Nebular properties}
In this section we use our GMOS spectroscopy to derive the nebular
properties of H\,{\sc ii} regions in NGC~5068, comparing our results
with previous estimates. Approximately 30\% of our spectroscopy
of WR candidates contained nebular lines in addition to the 4 H\,{\sc ii} regions that were
specifically targeted for this purpose. We performed Gaussian fits to
the excited nebular (and stellar) emission lines using the Emission
Line Fitting (\textsc{elf}) routine within the \textsc{starlink}
package \textsc{dipso}.

\subsection{Interstellar Extinction}
The interstellar extinction for each source was determined from the
observed Balmer line ratios of F$_{H\alpha}$/F$_{H\beta}$ with Case B
recombination theory \citep{Hummer1987}, assuming
n$_{e}$$\sim$100~cm$^{-3}$, T$_{e}$$\sim$10$^{4}$~K and a standard
Galactic extinction law \citep{Seaton1979}.

\begin{figure}
\includegraphics[width=0.73\columnwidth, angle=-90]{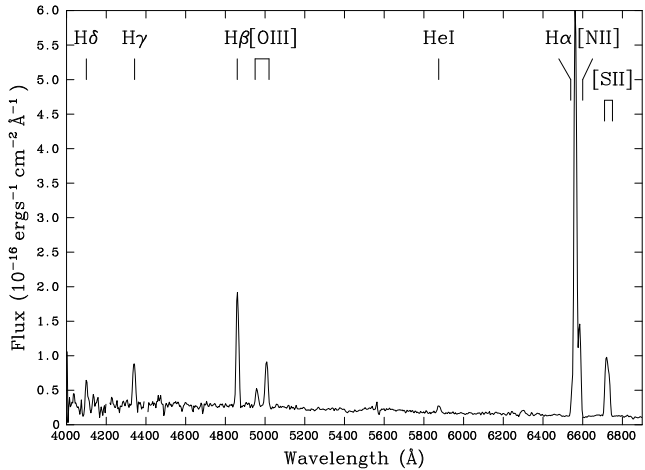}
\caption{The dereddened GMOS spectra of the source \#202, a H\,{\sc
    ii} region within NGC~5068. The emission lines used for nebular
  analysis are indicated. The extinction determined from this spectrum
  is E(B-V)~=~0.258\,mag, close to the average extinction.}
\label{nebular}
\end{figure}

Individual values of E(B--V) are listed for each source in Table
\ref{sources} and range from just above the foreground extinction at
E(B--V)\,=\,0.087\,mag to 0.428\,mag. The average extinction of
E(B--V)\,=\,0.259$\pm$0.035\,mag is applied to all sources for which
no spectroscopic value can be derived. Taking into account the
underlying Balmer absorption we estimate that the H$\beta$ flux is
underestimated by 1--13\% based on measured equivalent widths of
15-300~$\AA$.~\citet{Ryder1995} derive similar values of extinction
ranging from foreground extinction of E(B--V)\,=\,0.007\,mag to
E(B-V)\,=\,0.623\,mag. The dereddened emission line spectrum of the
H\,{\sc ii} regions \#202 is shown as an example in Figure
\ref{nebular} with all the of the nebular diagnostic lines indicated.

\subsection{Metallicity of NGC~5068}

Nebular emission lines from H\,{\sc ii} regions allow metallicity calculations
for NGC~5068. Previously, \citet{Ryder1995} used a
combination of long-slit spectroscopy and multi-aperture plate
slitlets to obtain spectra of 20 H\,{\sc ii} regions within
NGC~5068. Using the R$_{23}$ method \citep{Pagel1981} they determined
a metallicity gradient of
log(O/H)$+$12\,=\,(8.96$\pm$0.12)--(0.35$\pm$0.26)(r/R$_{25}$), where
r/R$_{25}$ is the deprojected distance from the centre of the galaxy
based on R$_{25}$$=$3.62\,arcmin for NGC~5068 from
\citet{deVau1991}. \citet{Pilyugin2004} subsequently combined the
spectroscopic observations by \citet{Ryder1995} with those from
\citet{McCall1985} and re-calculated a metallicity gradient of
log(O/H)$+$12\,=\,8.32$+$0.08(r/R$_{25}$) using the excitation
parameter, $P$, rather than from empirical values. This result 
suggests that the centre of the galaxy is more metal-poor than the outer
disk. Only 4 instances of positive gradients were found by
\citet{Pilyugin2004} from a sample of 54 galaxies. Usually,
spiral galaxies have flat or negative metallicity gradients \citep{Zaritsky1994},
e.g. NGC~7793 \citep{bc10} and M101\citep{Cedres2004}.

\begin{figure}
\centering
\includegraphics[width=1\columnwidth]{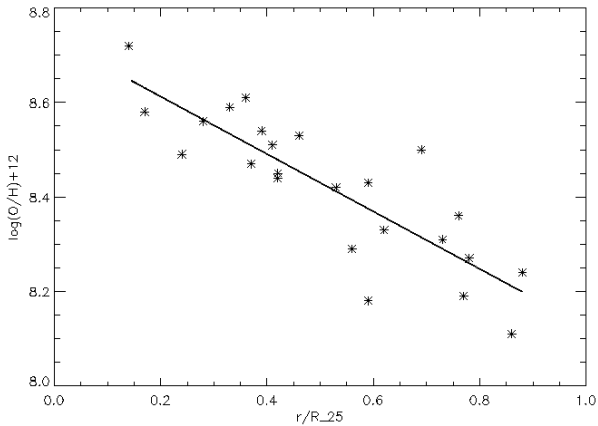}
\caption{Deprojected
  distance from the centre of the galaxy versus the metallicity
  derived for 21 H\,{\sc ii} regions in NGC~5068. The error on
  the individual points is $\pm$0.19.}
\label{gradient}
\end{figure}

Unfortunately, due to the low S/N of our spectroscopic observations
the $\lambda$4363$\AA$ [O\,{\sc iii}] line is not detected so an
accurate temperature, and hence emissivities cannot be
calculated. Consequently we are unable to use the weak-lined methods
described by \citet{Osterbrock1989} to determine the ionic abundances
of H\,{\sc ii} regions in NGC~5068; consequently we revert to
strong-line methods described by \citet{pp04}. In addition to the four
H\,{\sc ii} regions included in our MOS masks, the H$\alpha$, [N\,{\sc
  ii}], H$\beta$ and [O\,{\sc iii}] lines required are present in 17
WR spectra. We calculated the [N\,{\sc ii}]/H$\alpha$ (or N2) and
([O\,{\sc iii}]/H$\beta$)/N2 (or O3N2) line ratios by fitting the
nebular emission lines with a Gaussian profile using the Emission Line
Fitting routine (\textsc{elf}). The linear relationships of N2 and
O3N2 with metallicity from \citet{pp04} were used to determine the
metallicity of the H\,{\sc ii} region, shown in Table
\ref{abundances}. Using the gradient function in IDL we fit
the average metallicities of the H\,{\sc ii} regions and found a
metallicity gradient of
\begin{center}
\begin{displaymath}\log\frac{O}{H} + 12 = (8.74\pm0.15) - (0.61\pm0.22)\frac{r}{R_{25}} \end{displaymath}
\end{center}
for NGC~5068 which is shown in Figure \ref{gradient}.  Metallicities
range from log(O/H)$+$12\,=\,8.10 -- 8.72 with a central metallicity
of log(O/H)$+$12\,=\,8.74. The central metallicity is lower than that
found by \citet{Ryder1995}, however the metallicity gradient is
consistent within errors indicating that different methods used
(R$_{23}$ versus N2 and O3N2) to derive metallicity may have a
systematic offset \citep{Kewley2002,
    Kewley2008}. Unfortunately our spectral range does not include
the [O\,{\sc ii}] $\lambda$3727$\AA$ line required for the R$_{23}$
method, so a full comparison of the two methods cannot be
made. Finally, our negative metallicity gradient is inconsistent with
the positive gradient found by \citet{Pilyugin2004} and their estimate
of the central metallicity is $\sim$0.4\,dex lower than our own.

%This is
%supported by \citet{Garnett2004} who found that the oxygen abundance
%in metal-rich regions was overestimated using the R$_{23}$
%method. 

\begin{table}
\centering
\caption{N2 and O3N2 line ratios and corresponding metallicities for
  nebular sources in NGC~5068. The errors on log(O/H)+12 are
    dominated by the error on the fits to N2 ($\pm$0.41) and O3N2
    ($\pm$0.25) from \citet{pp04}. This corresponds to a error on the mean log(O/H)+12 of $\pm$0.24, and
    on the average log(O/H)+12 of $\pm$0.17.}

\begin{tabular}{c@{\hspace{2mm}}c@{\hspace{3mm}}c@{\hspace{3mm}}c@{\hspace{3mm}}c@{\hspace{3mm}}c@{\hspace{3mm}}c}
\hline\hline
Source & r/R$_{25}$ & \underline{I([N\,{\sc ii}])} & log(O/H) & \underline{I([O\,{\sc iii}])} & log(O/H) & log(O/H) \\
   ID  &           &   I(H$\alpha$)         & + 12 $^{1}$ &         I(H$\beta$)   & + 12 $^{2}$   & + 12 $_{mean}$ \\   
\hline
10  & 0.78 & 0.07 & 8.26 & 1.53 & 8.27 & 8.27\\
12  & 0.76 & 0.10 & 8.32 & 1.40 & 8.39 & 8.34 \\
21  & 0.77 & 0.07 & 8.25 & 7.06 & 8.13 & 8.19 \\
45  & 0.73 & 0.07 & 8.24 & 1.13 & 8.38 & 8.31 \\
H2  & 0.37 & 0.13 & 8.39 & 0.47 & 8.58 & 8.49 \\
54  & 0.53 & 0.15 & 8.42 & --   & --   & 8.42 \\
55  & 0.42 & 0.13 & 8.40 & 0.90 & 8.50 & 8.45 \\
60  & 0.41 & 0.12 & 8.37 & 0.26 & 8.65 & 8.51 \\
73  & 0.59 & 0.10 & 8.32 & 0.49 & 8.54 & 8.43 \\
99  & 0.33 & 0.44 & 8.70 & 0.93 & 8.47 & 8.58 \\
103   & 0.14 & 0.36 & 8.65 & 0.29 & 8.79 & 8.72 \\
104   & 0.09 & 0.37 & 8.65 & 0.19 & 8.43 & 8.54 \\
128   & 0.56 & 0.09 & 8.31 & 1.63 & 8.27 & 8.29 \\
129   & 0.42 & 0.20 & 8.50 & 3.22 & 8.38 & 8.44 \\
130   & 0.46 & 0.17 & 8.47 & 0.62 & 8.58 & 8.53 \\
131   & 0.17 & 0.26 & 8.56 & 0.55 & 8.66 & 8.61 \\
H3 & 0.69 & 0.14 & 8.41 & 0.22 & 8.70 & 8.56 \\
145 & 0.24 & 0.19 & 8.49 & --   & --   & 8.49 \\
H1 & 0.28 & 0.21 & 8.51 & 0.31 & 8.66 & 8.59 \\
149 & 0.86 & 0.04 & 8.10 & 4.20 & 8.11 & 8.10 \\
150 & 0.88 & 0.07 & 8.23 & 2.71 & 8.25 & 8.24 \\
\hline
Average &     &  0.155    &      &      &      & 8.44$\pm$0.17 \\
\hline
\multicolumn{7}{l}{$^{1}$ \footnotesize{N2}} \\
\multicolumn{7}{l}{$^{2}$ \footnotesize{O3N2}} \\ 

\end{tabular}
\label{abundances}
\end{table}

\section{Wolf-Rayet Population} \label{wr}

We have identified broad WR emission features in 30 of the 64
sources spectroscopically observed, i.e. only 47\% of these candidates
are confirmed. Note that these statistics exclude three candidates for
which slits were misaligned, as a result of conversion between the
FORS1 $\lambda$4684 imaging to the g' GMOS pre-imaging for the MOS
mask design.

Of the remaining spectroscopically observed candidates lacking
WR emission features, one is a early B-type star 
while the others show no stellar
features at the low S/N of the observations. 
Photometrically, the majority of these sources indicated
negligible or weak He\,{\sc ii}$\lambda$4686 excesses.  Indeed, 
only one (\#136) is a firm WR candidate with a $\lambda$4684 excess of 
$\geq$3$\sigma$, such that 21 of the 22 statistically
significant sources were confirmed as Wolf-Rayet stars.

Emission line fluxes of He\,{\sc ii}$\lambda$4686\,$+$\,N\,{\sc
  iii-v}$\lambda$4603-4641 for WN stars and C\,{\sc
  iii}$\lambda$4647-4651\,$+$\,C\,{\sc iv}$\lambda$5808 for WC stars,
are fit with Gaussian profiles using the \textsc{elf} suite within the
\textsc{dipso} package.  Spectral classification systems by
\citet{Smith1996} and \citet{Crowther1998} are used to assign either a
WN or WC subtype, respectively. At a distance of 5.45\,Mpc we cannot
spatially resolve individual WR stars. We therefore adopt the luminosity
calibrations derived by \citet{Crowther2006}, which are based on
average line luminosities for single WR stars in the Large Magellanic
Cloud (LMC), to determine the number of WN and WC stars in each
source. LMC templates are used in view of the metallicity of NGC 5068
obtained in Section 3.  Table \ref{wr_table} records the line fluxes
and luminosities for each WR source, the classification assigned and
the the number of WR stars present in that region.

%\onecolumn

\begin{table*}
\centering
\caption[Observed fluxes and dereddened luminosities of WR features of
sources in NGC~5068]{WR features in NGC~5068. Observed flux
  (F$_{\lambda}$) and extinction corrected luminosities
  (L$_{\lambda}$) based on a distance of 5.45\,Mpc. Values have been corrected for slit loss. Values
  in parentheses indicates a less secure detection
  ($<$3$\sigma$). Number of WR stars are based on the line
  luminosities for one WR star from \citealt{Crowther2006}a, while
  N(O7\,{\sc v}) stars is based on L(H$\alpha$).} 
\begin{tabular}{c@{\hspace{2mm}}c@{\hspace{2mm}}c@{\hspace{2mm}}c@{\hspace{2mm}}c@{\hspace{2mm}}c@{\hspace{2mm}}c@{\hspace{2mm}}c@{\hspace{2mm}}c@{\hspace{2mm}}c@{\hspace{2mm}}c@{\hspace{2mm}}c}
\hline\hline
      &  & \multicolumn{6}{c}{F$_{\lambda}$($\times 10^{-16}$ erg          s$^{-1}$ cm$^{-2}$)}  &  \multicolumn{2}{c}{L$_{\lambda}$($\times 10^{36}$ erg s$^{-1}$ )} &   \\ 
ID  & E(B-V) & F(N\,{\sc v-iii}) & F(C\,{\sc iii})   & F(He\,{\sc ii})  & F(He\,{\sc ii})  & F(C\,{\sc iii})  & F(C\,{\sc iv})  & L(He\,{\sc ii}) &  L(C\,{\sc iv}) & WR  &  N(WR)  \\
       & & 4603-4641    & 4647-4651     & 4686             &  5411        & 5696   & 5808     & 4686  &   5808    & Subtype &     \\  
\hline
6  & 0.247 & --            & --          & 3.51    & 0.30      & --    & --       & 2.96  &    --     & WN4--6 & 2  \\
10 & 0.428 & --            & 1.07        & --      & --        & --    & 1.62     & --    & 1.79      & WC4--5 & 1   \\
12  & 0.284 & --            & 2.04        & --      & --        & --    & 3.08     & --    & 2.32      & WCE    & 1  \\
13  & 0.465 & --            & --          & --      & --        & --    & 1.66     & --    & 2.02      & WCE(?) & 1   \\
17   & 0.230  & --            & 1.05        & --      & 2.04      & --    & 10.2     & --    & 6.67      & WC4--5 & 2   \\
19  & 0.259 & --            & 4.70        & --      & --        & --    & 5.05     & --    & 3.56      & WC     & 1  \\
21  & 0.595 & --            & --          & 1.36    & --        & --    & --       & 3.88  & --        & WNE(?) & 2  \\
35 & 0.259 & --            & 11.1        & --      & --        & --    & 8.53     & --    & 6.01      & WCE    & 2  \\
39  & 0.259 & --            & --          & 0.59    & --        & --    & --       & 0.52  & --        & WN     & 1  \\
43  & 0.259 & --            & --          & 5.00    & 1.59      & --    & 0.76     & 4.40  & 5.36      & WN5b   & 3    \\
59  & 0.259 & --            & 3.74        & --      & --        & --    & --       &  --   &  --       & WCpec?    & ?  \\
65  & 0.259 & --            & 14.9        & --      & --        & 0.70  & 9.29     & --    & 6.55      & WC6    & 2   \\
78  & 0.259 & --            & --          & 1.88    & --        & --    & --       & 1.65  & --        & WN6    & 1  \\
84  & 0.259 & --            & 12.8        & --      & --        & --    & 5.98     & --    & 4.22      & WC4--5 & 1  \\
90  & 0.259 & --            & 10.9        & --      & --        & --    & 8.86     & --    & 6.25      & WC4--5 & 2  \\
93  & 0.170 & --            & --          & 3.39    & --        & --    & --       & 2.18  & --        & WNE    &  1 \\
94  & 0.259 & --            & 8.67        & --      & --        & 1.23  & 3.48     & --    & 2.46      & WC7    & 1     \\
99  & 0.259 & 0.80          & --          & 5.36    & 0.99      & --    & --       & 4.62  & --        & WN6    & 3   \\
104 & 0.147 &  --           & --          & 3.86    & --        & --    & --       & 2.29  & --        & WN(?)  & 1   \\
106 & 0.087 & 0.56          & --          & 2.54    & --        & --    & --       & 1.22  & --        & WN7    & 1   \\
111  & 0.259 & --           & --          & 1.77    & --        & --    & --       & 1.56  & --        & WNL    & 1  \\
112  & 0.102 & --            & 1.62         & --      & --        & --   & 2.60     & --    & 1.21      & WC     & 1  \\
125  & 0.259 & --            & 12.1        & 1.53    & --        & --    & 5.85     & 1.34  & 4.12      &WC  & 1  \\
128  & 0.259 & --            & 4.59        & --      & --        & --    & 3.04     & --    & 2.14      & WC4--5 & 1  \\
131  & 0.180 & --            & 2.35        & 0.55    & --        & --    & 1.12     & 0.36 & 0.64      & WC4    & 1   \\
133  & 0.259 & --            & 5.67        & --      & --        & --    & 2.73     & --    & 1.93      & WC4    & 1   \\
134  & 0.181 & --            & 8.90        & --      & --        & --    & 19.3     & --    & 11.1     & WC4    & 3   \\
145  & 0.259 & --            & 0.85        & --      & --        & 0.69  & 4.12     & --    & 2.90      & WC6    & 1   \\
149   & 0.125 & --            & --          & 2.71    & --        & --    & 1.51     & 1.49  & 0.75      & WNE    & 1   \\
150   & 0.259 & --            & 9.81        & --      & --        & --    & 5.94     & --    & 4.19      & WC4--5 & 1  \\
151  & 0.259 & --            & --          & 1.63    & 0.24      & --    & 0.35     & 1.43  & 0.25      & WNE    & 1   \\
\hline

\end{tabular}
\label{wr_table}
\end{table*}

%\twocolumn

\subsection{WN stars}

Twelve of the 30 sources confirmed reveal the characteristic signature
of WN stars. If N\,{\sc v-iii}$\lambda$4603--4641 is detected then its
strength relative to the He\,{\sc ii}$\lambda$ 4686 emission line is
used to assign a refined classification of WN6 or WN7. For those WN
stars where N\,{\sc v-iii}$\lambda$4603--4641 was not detected we
assume a WNE subtype. Mid-type WN4--6 classifications are used when
both He\,{\sc ii}$\lambda$4686 and He\,{\sc ii}$\lambda$5411 are
detected. In one case we classify the WR source as WN5b since the WR
emission features are much broader than typical WR stars.  In total,
we identify 18 WN stars in the 12 sources.  Figure
\ref{wr_spectra_5068}a) shows the observed WN spectra of source \#99,
which hosts two WN5-6 stars. The template spectra, shown by the dashed
line, are taken from \citet{Crowther2006}. The He\,{\sc
  ii}$\lambda$4686, $\lambda$5411 and N\,{\sc iii}$\lambda$4641 WR
emission lines are marked while the additional narrow lines are
nebular emission lines.

\subsection{WC stars}\label{spec_wc}

\begin{figure}
\centering
\subfigure{\includegraphics[width=0.7\columnwidth, angle=-90]{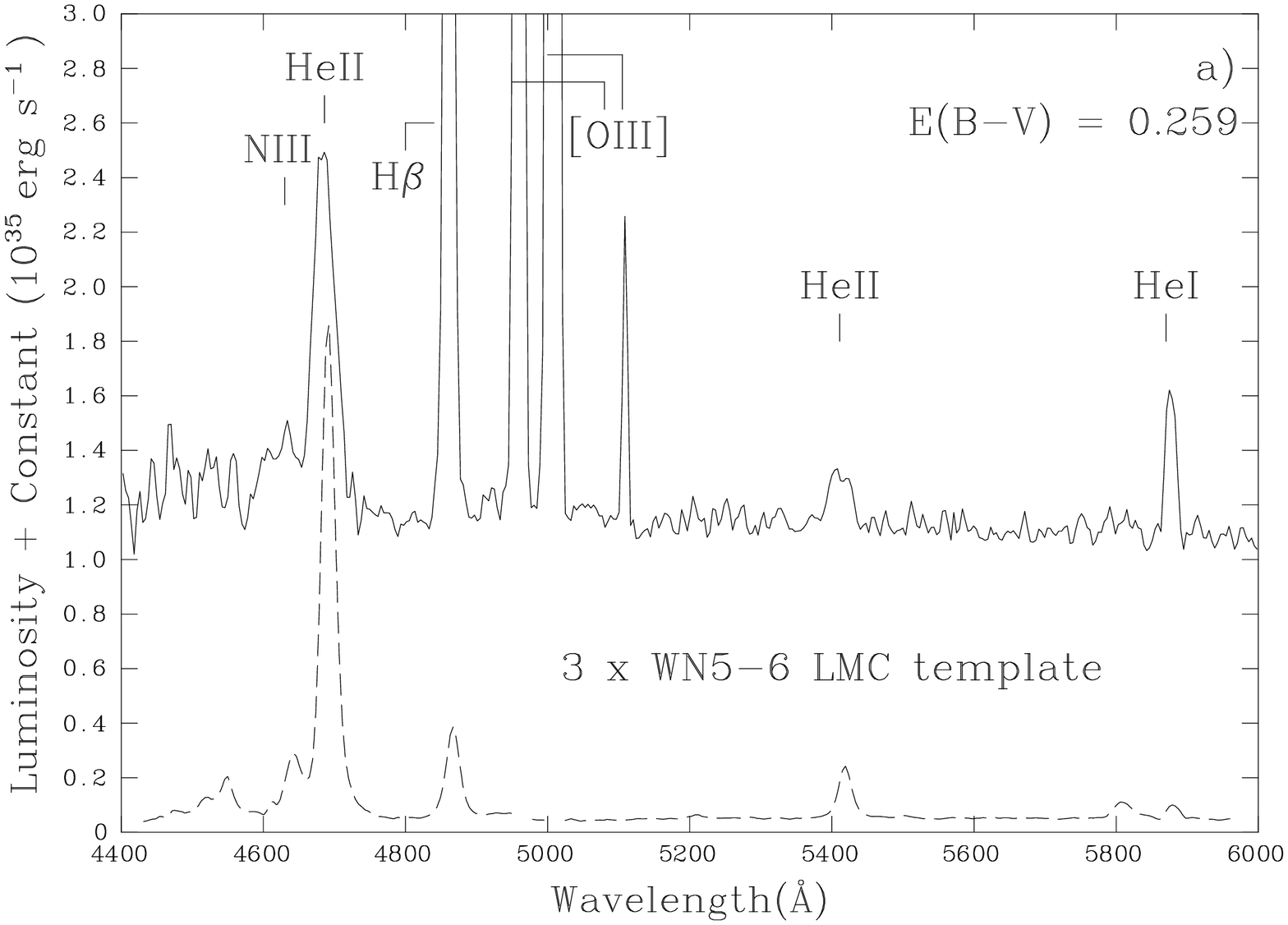}}
\subfigure{\includegraphics[width=0.7\columnwidth, angle=-90]{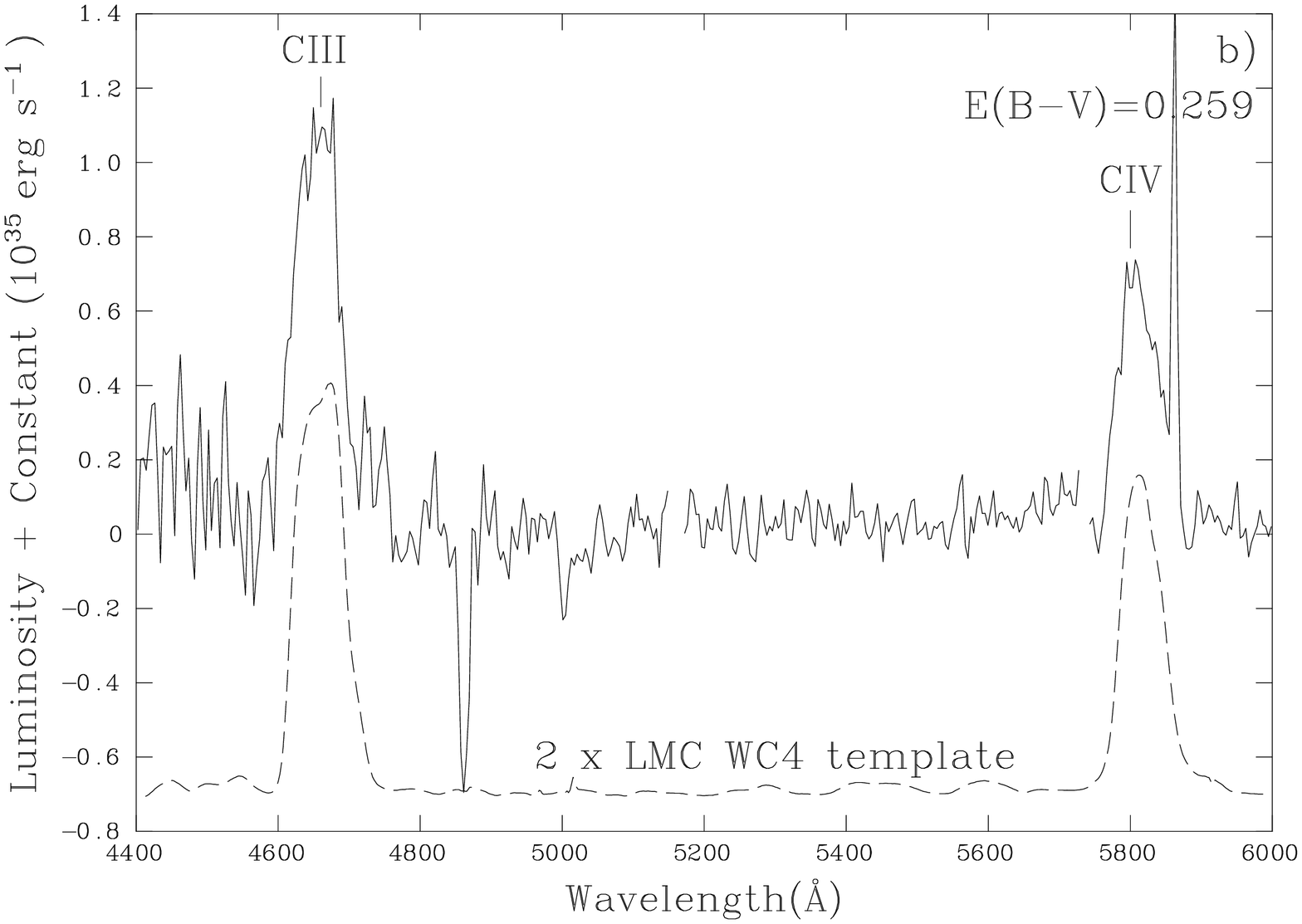}}
\caption{Dereddened spectra of a) WN (source \#99) and b) WC (source
  \#90) stars in NGC~5068 (solid lines), corrected for a distance of
  5.45Mpc, with LMC WR template spectra (dashed lines) offset for
  comparison. The WR emission lines (and nebular lines in a) are
  indicated and both sources have been extinction corrected by
  E(B-V)\,=\,0.259\,mag. The oversubtracted H$\beta$ line in b) is
  likely due to a bad sky subtraction.}
\label{wr_spectra_5068}
\end{figure}

We identify strong blue and red emission WC features for 18 of the 30
sources spectroscopically confirmed as WR stars.  Following the
classification of \citet{Crowther1998} we use the C\,{\sc
  iii}$\lambda$5696/C\,{\sc iv}$\lambda$5808 ratio to refine the
subtype of the WC stars. We find that WC4--5 stars dominate the WC
population, with no C\,{\sc iii}$\lambda$5696 detected in the stellar
spectrum, as is the case in other relatively metal-poor galaxies. A
typical example is presented in Figure \ref{wr_spectra_5068}b together
with an LMC WC4 template \citep{Crowther2006}. These are used to
determine the number of WC stars in each region, with most regions
hosting a single WC star. In total, we identify 24 WC stars in the 18
sources exhibiting WC emission.

%Source \#112 exhibits
%extremely strong carbon lines, although the C\,{\sc iii}$\lambda$4650
%and C\,{\sc iv}$\lambda$5808 cannot be matched simultaneously using the
%LMC templates. Figure \ref{wc_multiple} presents fits for a 
%range of extinctions (since this could not be
%measured directly) from which we conclude that only 
%foreground extinction of E(B--V)\,=\,0.102\,mag provides the optimal
%match, requiring 11 WC4 stars. Still, although
%C\,{\sc iv}$\lambda$5808 can be matched, C\,{\sc
%iii}$\lambda$4650 remains significantly underestimated. 

\begin{figure}
\centering
\includegraphics[width=0.7\columnwidth, angle=-90]{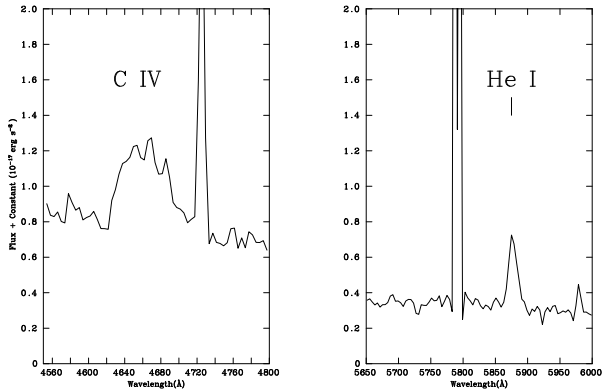}
\caption{Flux calibrated spectra of Source \#59 showing the presence
of broad $\lambda$4660 emission suggestive of C\,{\sc iv} $\lambda$4660
(left), albeit without neither C\,{\sc
iv}$\lambda$5808 nor C\,{\sc iii}$\lambda$5696 emission. 
The line at $\sim$5780$\AA$ is an unsubtracted sky line.}
\label{source_91_nociv}
\end{figure}

In addition to the 30 confirmed WR sources, broad emission at
$\lambda$4660 is seen in source \#59, which we attribute to either
C\,{\sc iii}$\lambda$4650, C\,{\sc iv} $\lambda$4660 or He\,{\sc ii}
$\lambda$4686, as indicated in Figure \ref{source_91_nociv}. However,
neither C\,{\sc iv}$\lambda$5808 nor C\,{\sc iii}$\lambda$5696 is
observed, from which we assign a WC?pec spectral type, but do not
attempt to quantify the number of WR stars.

%\begin{figure}
%\centering
%\includegraphics[width=0.7\columnwidth, angle=-90]{source_91.eps}
%\caption{Comparison of the WC source in NGC~5068 with LMC
 % WC templates of different populations. The solid line shows the
 % observed spectra source \#sp112 in a) the dashed line shows a fit of 17
 % WC4 stars and the dotted line shows a fit of 14 WC4 stars using an
 % extinction of E(B-V)=0.259. b) the dashed line in b shows a fit of 11
 % WC4 stars with E(B-V)=0.102.}
%\label{wc_multiple}
%\end{figure}

%\subsection{Composite Spectra}
% One WR candidate in NGC~5068 exhibits both strong He\,{\sc ii}$\lambda$4686 and C\,{\sc iv}$\lambda$5808 indicating the presence of both WN and WC stars. Figure \ref{composite} shows the best fit of combined templates LMC WR stars and the individual contributions from both the WN and WC stars to the calibrated spectra of source \#75; the region is found to host 2 WC4 and a single WN2--4 star.

%\begin{figure}
%\centering
%\includegraphics[width=0.70\columnwidth, angle=-90]{source_sp75.eps}
%\caption[Composite WN+WC spectra from NGC~5068 with LMC
 % templates]{Comparison of the WN+WC composite source in NGC~5068 with
 % LMC WC templates of different populations. The solid line shows the
 % observed spectra of source \#sp75 and the dashed line shows an the
 % combined LMC template for 2 x WC4 and 1 x WN2-4 stars. The dotted
 % line shows just the WC contribution and the dot-dashed line shows
 % the WN contribution offset from the observed spectra.}
%\label{composite}
%\end{figure}

\section{H$\alpha$ Imaging} \label{ha_section} 

We have obtained narrow-band $\lambda$6563 (H$\alpha$) and adjacent
continuum images using VLT/FORS1 to derive the total O star
population and SFR of NGC~5068. Images were absolutely flux calibrated
using the spectrophotometric standard star LTT~7987 \citep{Hamuy1994}
obtained on the same night as the science observations. The
$\lambda$6563 observations are contaminated by the [N\,{\sc
  ii}]$\lambda$6548,6583 doublet which can account for as much as
$\sim$30\% of the total emission in metal-rich regions.  Our
spectroscopic observations of H\,{\sc ii} regions resolve the
H$\alpha$ and [N\,{\sc ii}] lines finding on average [N\,{\sc
  ii}]/H$\alpha$\,=\,0.155 (Table \ref{abundances}) which we
applied to our measured $\lambda$6563 fluxes. We corrected the
H$\alpha$ fluxes for extinction using measured values where possible,
otherwise applying the average E(B--V)\,=\,0.259\,mag, before
calculating the final H$\alpha$ luminosities adopting a distance of
5.45\,Mpc \citep{Herrmann2008}.

\subsection{Star Formation Rate} \label{sfr_section} 

The star-formation rate of the galaxy was found by placing a circular
aperture with a diameter of 6.5~arcmin over the entire galaxy in
our VLT/FORS1 field of view using the \textsc{starlink} package
\textsc{gaia}. This observed H$\alpha$ flux was corrected for [N\,{\sc
ii}] emission ([N\,{\sc ii}]/H$\alpha$=0.155) and extinction
(E(B-V)=0.259\,mag) producing a total H$\alpha$ luminosity of
5.68$\times$10$^{40}$ erg s$^{-1}$. Following the relations of
\citet{Kennicutt1998} the corrected H$\alpha$ flux luminosity was used
to calculate the number of ionising photons, from which the SFR of the
galaxy is inferred. The VLT/FORS1 images of NGC~5068 suggest a
SFR\,=\,0.45\,M$_{\odot}$yr$^{-1}$, excluding the H$\alpha$ emission
which lies in the CCD chip gap and also extended emission beyond our field
of view.

Table \ref{sfr_5068} shows a comparison of our analysis with previous
work. The result from \citet{Ryder1994} is substantially higher than
the other estimates, but unfortunately no observed flux measurements
are available so no further comparison with this result can be
made. In addition, the observations of \citet{Hodge1974} are
consistent with our estimate, however, they only observe the inner
part of the galaxy and again no measured fluxes are available for
comparison.  We include both of these results in Table \ref{sfr_5068}
only for completeness.  The flux measurements of
\citet{Kennicutt2008} are the most complete since the
13~arcmin$^{2}$ field of view, which is double our own, contains all
the H$\alpha$ emission from the galaxy. The SFR they derive is
0.53\,M$_{\odot}$yr$^{-1}$, however they only correct for foreground
extinction, and find a larger [N\,{\sc ii}]/H$\alpha$ contribution of
0.209 which was estimated using empirical scaling relation since no
direct observations were available.

The H$\alpha$ flux from the archival CTIO images of NGC~5068
used by \citet{Kennicutt2008} exceeds our measurements by
$\sim$30\%, which is unsurprising considering the smaller field of
view and detector gaps in our VLT images.  Therefore, we adopt the
observed flux from \citet{Kennicutt2008} of F(H$\alpha$+[N\,{\sc
  ii}])\,=\,1.48$\times$10$^{-11}$erg\,s$^{-1}$\,cm$^{-2}$.
Since [NII]/H$\alpha$ decreases with metallicity from the
  metal--rich inner disk to the metal--poor outer disk, we correct for
  three separate values of [N\,{\sc ii}]/H$\alpha$ at r/ R$_{25}$\,=\,
  0--0.28, 0.29--0.69, 0.70--1.  Using the measured values of [N\,{\sc
    ii}]/H$\alpha$ from sources listed in Table \ref{abundances} we
  find the average [N\,{\sc ii}]/ H$\alpha$ value within each r/
  R$_{25}$ range, resulting in [N\,{\sc ii}]/ H$\alpha$\,=\,0.28,
  0.17, and 0.08, from the inner to outer disk, respectively.  We find
  a F(H$\alpha$)\,=\,1.22$\times$10$^{-11}$erg\,s$^{-1}$\,cm$^{-2}$
  and calculate total a H$\alpha$ luminosity of
7.93$^{+1.32}_{-1.65}$$\times$10$^{40}$ erg s$^{-1}$ for
NGC~5068, after correction for the [N\,{\sc ii}] emission and average
extinction from our GMOS spectroscopy. This corresponds to a
global SFR\,=\,{0.63$^{+0.11}_{-0.13}$\,M$_{\odot}$yr$^{-1}$
adopting the \citet{Kennicutt1998} calibration. The errors on
  the SFR are derived from the errors on E(B-V) and on the distance of
  NGC~5068 from \citet{Herrmann2008}.

\begin{table}

\centering
\caption{Integrated H$\alpha$ flux/luminosity for NGC~5068, with respect to 
\citet{Kennicutt2008} (K08) and other previous studies. Fluxes are in units of
  10$^{-11}$ erg\,s$^{-1}$\,cm$^{-2}$, and luminosities are in units of
  10$^{40}$ erg\,s$^{-1}$. The inner, middle and outer regions
  correspond to r/ R$_{25}$\,=\,0--0.28, 0.29--0.69, 0.7--1.0, respectively.}
\begin{tabular}{l@{\hspace{1mm}}c@{\hspace{2mm}}c@{\hspace{2mm}}c@{\hspace{1mm}}c@{\hspace{1mm}}c@{\hspace{1mm}}c@{\hspace{1mm}}c@{\hspace{-2mm}}c}
\hline\hline
Work & F(H$\alpha$ & \underline{[N\,{\sc ii}]} & E(B-V) & D  &  L(H$\alpha$) & log(Q$_{0}$) & N(O7\,{\sc v}) &  SFR  \\
       &      +[N\,{\sc ii}])  & H$\alpha$       &        & Mpc &            &  s$^{-1}$  &      & \footnotesize{M$_{\odot}$\,yr$^{-1}$} \\              
\hline
This work  & 1.04 & 0.155 & 0.259 & 5.45 & 5.68 & 52.62 & 4200 & 0.45 \\
K08 & 1.48 & 0.209 & 0.090  & 6.20 & 6.61 & 52.69 & 4900 & 0.53 \\
% & 1.48 & 0.155 & 0.259 & 5.45 & 8.09 & 52.77 & 6000 & 0.65 \\
\hspace{4mm}Inner$^{1}$ & 0.23  & 0.28 & 0.259 & 5.45 & 1.07 & 51.90  & 800 & 0.09 \\
\hspace{4mm}Middle$^{2}$ & 1.04 & 0.18 & 0.259 & 5.45 & 5.53 & 52.62 & 4200 & 0.45 \\
\hspace{4mm}Outer$^{3}$ & 0.21   & 0.08 & 0.259 & 5.45 & 1.26 & 51.97 & 900 & 0.10 \\
Total & 1.48  &          &           &         & 7.93 & 52.77 & 5900 & 0.63 \\
\multicolumn{8}{l}{\citet{Ryder1994}} & 3.0 \\
\multicolumn{8}{l}{\citet{Hodge1974}} & 0.35 \\
\hline
\end{tabular}
\label{sfr_5068}
\end{table}

\subsection{O Star Population} \label{O_pop}

The H$\alpha$ luminosity of
7.93$^{+1.32}_{-1.65}$$\times$10$^{40}$ erg s$^{-1}$
equates to $\sim$6000$^{+2000}_{-1500}$ O7\,{\sc v} stars,
assuming an ionising flux Q$_{0}$\,=\,10$^{49}$ ph\,s$^{-1}$ for a
typical O7\,{\sc v} star in a low metallicity environment
\citep{Vacca1994}. However, in reality not all of the O stars will be
O7 dwarfs but will span the entire spectral range. Starburst models by
\citet{Schaerer1998} can be used to estimate the true number of O
stars for a given age, $t$, using the parameter
$\eta_0(t)$. Unfortunately the spiral arms of NGC~5068 undergo
continuous star-formation and hence the age could range from
3--10\,Myr, the lifetime of O stars. Consequently we follow
\citet{bc10} who used the 30 Doradus region of the LMC, and the SMC,
as proxy to determine a uniform correction factor of N(O)/N(O7\,{\sc
  v})$\sim$1.5, suggesting a total O star population of
$\sim$9000$^{+1500}_{-2000}$.

\citet{Hodge1974} identified 83 H\,{\sc ii} regions in NGC~5068,
including some giant H\,{\sc ii} regions (GHR), formally defined 
as regions in which Q$_{0}$ exceeds 10$^{50}$ photon~s$^{-1}$ \citep{Conti2008}.
A catalogue of the 29 brightest H\,{\sc ii} regions in NGC~5068 is
presented in Table \ref{ha}. 
The observed H$\alpha$ flux within a given
diameter, d$_{\rm ap}$, has been continuum subtracted, corrected for
the contribution of [N\,{\sc ii}], plus an average NGC~5068 extinction
unless this has been individually measured (see Section \ref{ha_section}). 
From the list of H\,{\sc ii} regions in Table~\ref{ha}, $\sim$17 qualify
as GHR, albeit these are dependent upon extinction and distance 
uncertainties. Note that  NGC~5068 hosts three
\textit{very} bright H\,{\sc ii} regions, which each host over 100 equivalent
O7 dwarfs, reminiscent of the brightest H\,{\sc ii} regions in the Milky 
Way (e.g. NGC~3603), albeit somewhat more modest in ionizing output than the 
30 Doradus region of the LMC \citep{Kennicutt1984}.

Mindful that Table \ref{ha} may be incomplete due to the
detector gap in the FORS1 images, we inspected
the archival H$\alpha$ CTIO images of \citet{Kennicutt2008}
but found that no substantial H\,{\sc ii} regions were present
in this region.

Table \ref{ha} provides cross matches to the catalogue of \citet{Hodge1974}, and 
also indicates which bright H\,{\sc ii} regions host WR candidates. 20\% of the
WR candidates lie within the most prominent H\,{\sc ii} regions, although considering 
the entire 160 candidate list, 50\% lie within bright (compact) 
H\,{\sc ii} regions, 25\% are associated with faint (diffuse) nebulosity, and 
25\% are not associated with any H\,{\sc ii} region (Table~\ref{sources}).

\begin{table*}
\centering
\caption{The flux,
  luminosity and N(O) stars for the brightest 29 H\,{\sc ii} regions in
  NGC~5068. Continuum subtracted fluxes, F(H$\alpha$), are in units of
  10$^{-14}$ erg\,s$^{-1}$\,cm$^{-2}$ and are corrected for [N\,{\sc
    ii}] emission using H$\alpha$/[N\,{\sc ii}]\,=\,0.155.
  Luminosities are in units of 10$^{38}$ erg\,s$^{-1}$, assuming a
  distance of 5.45\,Mpc. The diameter of the aperture, d$_{\rm ap}$ is given in
  arcsec. Also noted is whether WR stars have been identified within
  these H\,{\sc ii} regions.}
\begin{tabular}{lcccccccccc}
\hline\hline
ID	& RA & Dec &  d$_{\rm ap}$ & F(H$\alpha$)	& E(B-V)      &
L(H$\alpha$) & log(Q$_{0}$)	& N(O7\,{\sc v})  & Hodge & WR \\
\hline
\hline
HII \#1 & 13:18:42.8 & -21:01:24 & 5 & 19.1 & 0.259 & 12.3 & 50.96 & 91 & H82 & --   \\ 
HII \#2 & 13:18:44.5 & -21:01:14 & 5 & 1.72 & 0.259 & 1.11 & 49.92 & 8 & H79 & -- \\
HII \#3 & 13:18:44.8 & -21:00:51 & 4 & 1.48 & 0.259 & 0.96 & 49.85 & 7 & H77 & 11 \\
HII \#4 & 13:18:45.8 & -21:02:31 & 5 & 4.28 & 0.259 & 2.77 & 50.31 & 20 & H73 & 14,16\\
HII \#5 & 13:18:47.0 & -21:00:50 & 5 & 1.69 & 0.259 & 1.09 & 49.91 & 8 & H86 & -- \\
HII \#6  & 13:18:47.1 & -21:00:34 & 3 & 0.76 & 0.595 & 1.07 & 49.90 & 9 & & 21 \\
HII \#7 & 13:18:47.3 & -21:03:53 & 6 & 3.18 & 0.259 & 2.05 &50.18 & 15 && 23,26 \\
HII \#8 & 13:18:48.0 & -21:03:11 & 4 & 1.63 & 0.259 & 1.05 & 49.89 & 8 &  H65 & -- \\
HII \#9 & 13:18:48.1 & -21:00:49 & 7 & 16.9  & 0.259 & 10.9 & 50.91 & 81 &H60 & 34,36 \\
HII \#10 & 13:18:48.5 & -21:00:17 & 4 & 1.75 & 0.259 & 1.13 & 49.92 &8 & H59 & 40 \\
HII \#11 & 13:18:49.0 & -21:00:22 & 4 & 4.08 & 0.259 & 2.64 & 50.29 & 20 & H54  & 41 \\
HII \#12 & 13:18:49.3 & -21:00:23 & 4 & 4.98 & 0.259 & 3.22 & 50.38 & 24 & H50 & 45 \\
HII \#13 & 13:18:49.8 & -21:02:29 & 5 & 4.28 & 0.259 & 2.77 & 50.31 & 20 & H48 &  52,53,56 \\
HII \#14 & 13:18:49.9 & -21:03:23 & 10& 28.2 & 0.259 & 18.3  & 51.13 & 135 & H49 & 49,55,58,60 \\
HII \#15 & 13:18:50.7 & -21:02:06 & 5 & 1.39 & 0.259 & 0.90 & 49.82 & 7 & H42 & 69,70 \\
HII \#16 & 13:18:51.3 & -21:04:09 & 5 & 3.22 & 0.259 & 2.08 & 50.19 & 15 & H40 & 73,75,76 \\
HII \#17  & 13:18:51.7 & -21:01:18 & 5 & 7.16 & 0.259 & 4.63 & 50.53 & 34 & 78 \\
HII \#18  & 13:18:51.8 & -21:00:50 & 4 & 1.86 & 0.259 & 1.20 & 49.95 & 9 & H36 & 81 \\
%GHII \#19  & 13:18:51.9 & -21:01:16 & 6 & 50.0 & 0.259 & 32.4 & 50.38 & & H38 & 82,85 \\
HII \#20  & 13:18:53.2 & -21:02:45 & 4 & 2.72 & 0.259 & 1.76 & 50.12 & 13 & H33 & 103 \\
HII \#21 & 13:18:54.7 & -21:03:10 & 7 & 4.34 & 0.259 & 2.81 & 50.32 & 21 & H29 & 116 \\
HII \#22 & 13:18:54.9 & -21:04:04 & 4 & 2.94 & 0.259 & 1.32 & 49.99 & 10 &  & 121\\
HII \#23 & 13:18:55.2 & -21:02:42 & 4 & 1.95 & 0.259 & 1.26 & 49.97 & 9 & & -- \\
HII \#24 & 13:18:56.7 & -21:00:33 & 4 & 1.53 & 0.259 & 9.87 & 49.86 & 7 & H14 & -- \\
HII \#25 & 13:18:56.8 & -21:04:13 & 4 & 5.18 & 0.181 & 2.79 & 50.32 & 21 & & 134 \\
%GHII \#26 & 13:18:56.9 & -21:04:13 & 6 & 60.0 & 0.259  & 38.9 & 50.46 & & H16 & 138 \\
HII \#27  & 13:18:57.3 & -21:00:47 & 15 & 29.1 & 0.259 & 18.8 & 51.14 & 139 & H13 & 139 \\
HII \#28  & 13:18:58.2 & -21:02:28 & 4 & 2.72 & 0.259 & 1.76 & 50.11 &13  & & 145 \\
%GHII \#29  & 13:18:58.2 & -21:02:31 & 4 & 25.0 & 0.259 & 16.2 & 50.08 & & H12 & -- \\
HII \#30 & 13:18:59.2 & -21:00:06 & 4 & 1.20 & 0.259 & 0.78 & 49.76 & 6 & H10 & -- \\
HII \#31  & 13:19:00.8 & -21:02:47 & 9 & 3.21 & 0.259 & 20.8 & 51.19 & 154 & H7 & 152 \\
HII \#32 & 13:19:03.1 & -21:01:00 & 4 & 1.51 & 0.259 & 0.98 & 49.86 & 7 & H2 & -- \\
%\vspace{0.005mm} \\
%GH\,{\sc ii} \#1  & 8.4   & 352.3 & 0.259 & 228.1 & 51.23 & 169   \\
 % \multicolumn{7}{l}{(13:19:00.735   --21:02:47.62)}  \\
%GH\,{\sc ii} \#2  & 10.8  & 303.5 & 0.259 & 196.5 & 51.16 & 145 \\
%\multicolumn{7}{l}{(13:18:57.184   --21:00:46.35)} \\
%GH\,{\sc ii} \#3  & 8.3   & 350.5 & 0.259 & 226.9 & 51.22 & 168  \\
%\multicolumn{7}{l}{(13:18:49.920   --21:03:23.01)}  \\
\hline
\end{tabular}
\label{ha}
\end{table*}

\section{The Global WR Population of NGC~5068}

The spectroscopic observations of 30 WR candidates identifies 18 WN stars
and 24 WC stars in NGC~5068. 
We use the photometric properties of these confirmed WR
stars to infer the likelihood that the remaining WR candidates are
indeed true WR stars, and in some cases assess their WR subtype. In
addition we take account of the detection limits and photometric
completeness of our survey to derive the global WR population of NGC~5068.

\subsection{Nature of the Remaining Candidates}

%Firstly, we need to ascertain whether the photometric properties of
%the WR stars are consistent with the observed spectrum. Spectroscopic
%magnitudes were derived by convolving the observed flux calibrated
%object spectrum with the filter bandpass through which the photometry
%was obtained. The results of this comparison are shown in Figure
%\ref{outliers} and indicate that the photometric magnitudes are
%consistent with the spectroscopic magnitudes. The only outlier is
%source \# 17 which is spectroscopically shown to host a WC star,
%however photometry indicated no $\lambda$4684 excess suggesting the
%photometry of this region is unreliable. This source was included as a
%candidate following visual inspection of the continuum subtracted
%image which clearly showed the presence of$\lambda$4684 emission.

In section \ref{MOS} we established that our photometric and
spectroscopic $\lambda$4684 excesses were in good agreement 
(recall Figure \ref{outliers}). Therefore, we can assess
the nature of the remaining candidates based upon the properties of
our confirmed WR sources. 
Figure \ref{candidates_plot} compares the photometric m$_{4684}$--m$_{4781}$
excess against the absolute $\lambda$4684 magnitude of the WR
candidates identified in the VLT/FORS1 imaging. Those candidates, which
have a spectroscopically determined subtype are plotted accordingly,
while those candidates for which spectroscopy was not obtained are
plotted by the open triangles and the non-WR sources are marked with
an ``x".

In addition to the sources plotted in Figure \ref{candidates_plot}
there are 3 confirmed WC, 2 confirmed WN sources and 25 additional
non-WR sources which do not possess photometry in one or both
filters. Photometry for such sources is not available if the source is extended, in a crowded 
region, or too faint in the $\lambda$4781 observation.

As a result of their strong emission lines, 
WC stars have significantly larger photometric $\lambda$4684 excesses relative to WN
stars\citep{Conti1989}.  Our spectroscopic observations 
reflect this, since all sources with
$\lambda$4686 excesses of at least --1.5\,mag 
are spectroscopically confirmed as WC stars, as indicated in Figure
\ref{candidates_plot}. Consequently, we can fairly reliably 
infer that the remaining 5
WR candidate regions with equally large $\lambda$4684 excesses also host WC
stars.

In order to estimate the number of Wolf-Rayet stars for each
photometric candidate, we convert $m_{\rm 4684}$, and $m_{\rm 4781}$
magnitudes into fluxes, and estimate emission line fluxes by
subtracting $f_{\rm 4781}$ from $f_{\rm 4684}$, and then follow the
same approach as for the spectroscopic datasets. We have compared
photometric and spectroscopic 4603/4650/4686 fluxes for 24 sources in
common, and find spectroscopic fluxes are typically $\sim$50\%
higher. In part, this arises from the relatively narrow FWHM of the
$\lambda$4684 bandpass with respect to broad WR emission
(e.g. FWHM$\sim$55$\AA$~ for \#112). Therefore, photometric flux
estimates should provide useful lower limits to the WR population of
each source. 

For the spectroscopically observed candidates with photometric
excesses between --0.6 and --1.5\,mag, all are confirmed except in one
instance, so these too are likely to host WR stars.
It is less straightforward to assign a WN or WC subtype to these 
candidates since they overlap in $\lambda$4684 excesses since 
the stellar continua from other stars dilute the WR emission lines, resulting in
smaller excesses; this is discussed further in Section
\ref{completeness}. Still, since WN subtypes possess intrinsically
weak emission, they are more likely to correspond to the faintest sources with 
modest excesses. 

%For candidates with a $\lambda$4684 excess between 0 and --0.5\,mag,
%only $\sim$20\% of the spectroscopically observed candidates revealed
%WR features. Within this boundary there are a remaining 71 candidates
%and hence we assume only 14 of them are true WR stars. We note that
%this is a conservative estimate since fainter sources, with say
%M$_{4684}$$>$--6\,mag with m$_{4684}$--m$_{4781}$\,=\,--0.5\,mag are
%more likely to be WN stars which have weaker emission, however this
%region is not sampled in our spectroscopy so we withhold from making
%further conclusions.

Therefore, we assign a photometric WN subtype for sources fainter than
$M_{4686} \sim -7.5$, and photometric WC subtypes otherwise. For the
spectroscopic sources with weaker excesses below $-$0.6\,mag, only a
small percentage ($\sim$20\%) were confirmed as genuine WR
stars. Therefore, for such stars to be considered as a photometric WR
star we require a photometric m$_{4684}$--m$_{4781}$ excess of at
least 3$\sigma$, plus a excess in $\lambda$4686 with respect to the
$V$-band.  Such stars are most likely to be WN subtypes, except for
the very brightest sources with $M_{4686} \sim -8.5$ mag.

In total, we estimate an additional 43 WN stars to the 18
spectroscopically identified, plus an additional 11 WC stars to the 24
obtained from spectroscopy, i.e. 61 WN and 35 WC stars in total.
% 113

In addition to photometric sources, 17 candidates without photometric
information have yet to be incorporated. From the 7 such cases
spectroscopically observed, 5 were confirmed as WR stars, so we shall assume that 
a similar fraction (12 from 17) of the candidates are genuine
WR stars.
% 126
 However, we do not attempt to discriminate between WN or WC subtypes.
Our estimate for the grand total of WR stars in NGC 5068 is therefore
$\sim$110, with N(WN)/N(WC)$\geq$1.7.

%The fainter candidates
%(M$_{4684}$$>$--6\,mag) within the same $\lambda$4684 regime are
%another sample which was under--sampled in our follow-up spectroscopy
%and so we assume a notional 50\% WR confirmation rate to be cautious
%contributing an additional 25 WR stars to the WR population. However,
%the majority of WR candidates sit below our 100\% and 50\% detection
%limit so there are most likely WR stars present in NGC~5068 which our
%imaging (or resolution) was not sufficient enough to detect. The total
%number of WR stars we find is $\sim$114 WR stars.

 \begin{figure}
\centering
\includegraphics[width=0.74\columnwidth, angle=-90]{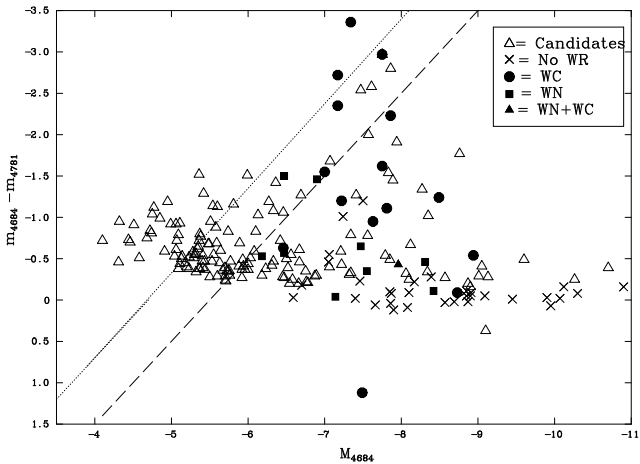}
\caption{Absolute
  He\,{\sc ii}\,$\lambda$\,4686 magnitude versus He\,{\sc
    ii}\,$\lambda$\,4686 excess for WR candidates in
  NGC~5068. Different symbols are used for WR subtypes confirmed by
  spectroscopy and are marked as shown in the legend. The 100\% and
  50\% completeness limits are indicated by the dashed and dotted
  lines, respectively.}
\label{candidates_plot}
\end{figure}

\subsection{Completeness} \label{detection_limits} \label{completeness}

From our 160 candidate WR sources 
in NGC~5068 we have estimated a global total of
$\sim$110 WR stars. 
However, in order to assess the true WR population we must assess how 
many WR stars were not detected due to the limiting magnitudes our 
images.

%account for our completeness of our observations and also our ability to detect WR stars in crowded regions. 

The detection limits plotted as the dashed (100\%) and solid (50\%)
lines in Figure \ref{candidates_plot} represent the depth of our FORS1
imaging based on the magnitude of all the objects detected in the
image (including non-WR sources). The $\lambda$4684 magnitude range of
objects was 16--27\,mag with the turnover of the distribution at
$\sim$24.0\,mag corresponding to a 100\% detection limit of
M$_{4684}$\,=--5.4\,mag for an average extinction of
E(B--V)\,=\,0.259\,mag and distance of 5.45\,Mpc.

%Accounting for the 50\% detection limits we would expect there to
%be an additional $\sim$10--20 WR stars in NGC~5068 that we have not
%detected, increasing the total number of WR stars to $\sim$120.

The main uncertainty on the assessing completeness of a global WR
population is identifying unresolved regions hosting WR stars. While
relatively isolated and resolved WR sources are easily identified from
narrow--band imaging, unresolved regions hosting multiple early-type stars
can dilute the WR emission and hide the WR source completely. How many
candidates we did not detect due to dilution?

\citet{bc10} derived the total completeness of their WR study of
NGC~7793 from comparison with the complete WR population of the LMC by
tracing how the $\lambda$4684 emission from a representative sample of
LMC WR stars would be diluted if the spatial resolution of the LMC
images were the same as their NGC~7793 imaging
(1.3\arcsec$\sim$20\,pc). By applying the detection limits of the
narrow-band imaging to the diluted LMC sample they estimated the
percentage of the known LMC WR stars that would not have been detected.
 
The larger distance of 5.45~Mpc to NGC~5068 together with an improved
spatial resolution of 0.8$''$ combine to produce a similar physical scale of 
$\sim$20\,pc. This is presented in Fig.~\ref{lmc_5068}, from which we
estimate that $\sim$85\%  of the WR stars in the LMC
would be detected in our survey, based on the detection
limits and spatial resolution of our narrow--band images, indicating a
global WR population of $\sim$130.

Overall, since WC stars have a stronger excess relative to WN stars it
is likely that we are close to $\sim$100\% complete in WC stars in
NGC~5068. Early--type WN stars (WNE) have larger excesses than
late--type WN stars (WNL), but are significantly fainter visually, so
we expect to be missing a significant percentage of WNE stars, but
also be relatively complete in WNL stars.  However, the diluted
photometric excesses of 4 out of the 15 representative WR stars were
$<$0.3 mag and fainter than $M_{4686}$ = --7 mag (Figure
~\ref{lmc_5068}.)  Some such faint, weak excess sources -- primarily
WNE stars -- will have been included in our catalogue, but would most
likely be listed under ``no WR'' since they are not photometrically
statistically significant. Indeed we see several examples of such
sources in our survey such as Source \#28, which is a genuine WNE star
with an excess of only 0.18\,mag and M$_{4684}$\,=\,--6.53\,mag. Our
survey of WNE stars is almost certainly incomplete from the
persepective that the large excesses, which enable us to identify
these faint stars, are diluted leaving only a faint excess which is
hard to detect from ground--based images. Therefore, we predict a
global WR population of $\sim$170 for NGC~5068 due to incompleteness
for the faintest 25\% of WR stars.

%However, 10 out of 15 of the representative WR stars had their
%excesses diluted to m$_{4684}$--m$_{4781}$$>$--0.5\,mag where we only
%had a 20\% success rate, hence we would only expect to detect two of
%the 10 sources in this excess range. Consequently as many as $\sim$50\% of
%the LMC WR stars would be missed from our survey. This can be used to
%place an upper limit on the number of WR stars in NGC~5068 of
%$\sim$240. 

%We must also ask ourselves, in what fraction of the LMC WR stars is
%the line dilution so extreme that the $\lambda$4684 excess no longer
%exists? Without obtaining high spatial resolution (ideally
%space--based) images of NGC~5068 a true evaluation of this effect is
%not possible. However, 

%None of our diluted LMC WR sample subtended the m$_{4684}$--m$_{4781}$\,=\,0 limit, however 

\begin{figure}
\centering
\includegraphics[width=0.73\columnwidth, angle=-90]{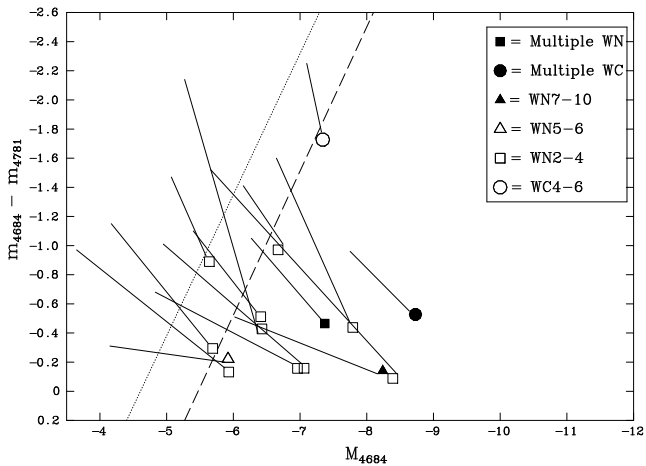}
\caption{Dilution of the
  $\lambda$4686 emission line for a representative sample of WR stars
  in the LMC, adapted from \citet{bc10}. We have applied our 100\% and 50\% 
  detection limits from
  our VLT/FORS narrow-band imaging, indicated by the dashed and dotted
  lines respectively, to assess the completeness of our WR survey of
  NGC 5068. 14 out of 15 LMC WR stars would have been detected
  indicating that our survey should be $\sim$90\% complete.}
%Dilution of the $\lambda$4684 emission of a
 % representative sample of WR stars in the LMC corresponding to a
 % spatial resolution of 0.8\arcsec or $\sim$20\,pc at a distance of
 % 4\,Mpc taken from \citet{bc10}. The dashed and dotted lines are the
 % 100\% and 50\% detection limits, respectively for the He\,{\sc
 %   ii}\,$\lambda$\,4684 image of NGC~5068. This indicates that
 % our survey would have detected $\sim$90\% of the WR stars if the LMC
 % were at 4\,Mpc. We note for the actual distance of NGC~5068 of
 % 5.45\,Mpc the sources would be shifted very slightly to the lower
 % right, however the difference would not make a difference to our
 % results.}
\label{lmc_5068}
\end{figure}

\subsection{Predicted WR population}

In Section \ref{O_pop} we used the SFR of the galaxy to determine that
there were $\sim$9000$^{+1000}_{-2000}$ O stars in
NGC~5068. We can use this information to infer the expected number of
WR stars based on empirical values and independently assess the
completeness of our survey.  The ratio of WR to O stars varies with
metallicity, as a result of the metallicity dependence of winds from
massive stars, ranging from $\sim$0.03 in the LMC to $\sim$0.15 in the
Solar neighbourhood \citep{Crowther2007}.  The average metallicity we
find for NGC~5068 is log(O/H)$+$12\,=\,8.44$\pm$0.17 (recall
  Table \ref{abundances}), similar to the LMC, so assuming a similar
WR to O ratio for NGC~5068 we would anticipate a WR population of
$\sim$270$^{+45}_{-60}$ based upon its O star population of
$\sim$9000$^{+1500}_{-2000}$.

If the WR population of NGC~5068 is $\sim$270, our earlier
estimate is 40\% incomplete. One possible explanation is that the
percentage of weak-lined, faint WN stars could be considerably higher
relative to other WR subtypes, as indeed is the case in the
LMC. Ideally, deep, high resolution (space--based) imaging of NGC~5068
would be required, together with further follow-up spectroscopy
to confirm this.
 
\section{Comparison to NGC~7793}

WR surveys beyond the local group have been undertaken on several
nearby, spiral galaxies (e.g. \cite{Crowther2006},
\cite{Hadfield2007}, \cite{bc10}). The survey of NGC~7793 by
\citet{bc10} is an ideal galaxy with which to compare NGC~5068 given
their similar physical properties.

NGC~7793 is a Sculptor group SA(s)d galaxy which is slightly nearer to
us at 3.91\,Mpc \citep{Karachentsev2003} than NGC~5068. Basic
properties of each galaxy are presented in Table \ref{gal_prop} which
shows that although the galaxies have similar physical sizes and
central metallicities, the abundance gradient of NGC~5068 is nearly
twice as steep as NGC~7793. We note that both studies use the
  N2 and O3N2 calibrations so are subject to the same systematic
  errors hence a comparison of H\,{\sc ii} region abundances is not
  biased. 

Also included in Table \ref{gal_prop} is a comparison of their star
formation rates, with NGC~5068 50\% higher than NGC~7793. A similar
number of WC stars have been spectroscopically confirmed in both
galaxies, and although a smaller fraction of WN have been observed in
NGC~5068 its WR population is anticipated to be $\sim$40\% higher.

\begin{table}
  \caption{Global properties of NGC~5068 and NGC~7793, along with a 
    census of their massive stellar populations. Properties of NGC~7793 are taken from \citet{bc10} unless otherwise stated. The number of WR candidates and the percentage that are spectroscopically confirmed are listed along with the confirmed number of WN and WC stars.} 
\begin{tabular}{ccc}
\hline
\hline
Name & NGC~7793 & NGC~5068 \\
\hline
Hubble Type & SA(s)d & SB(d) \\
Distance (Mpc) & 3.91$^{a}$ & 5.45$^{b}$ \\
R$_{25}$ (arcmin) & 4.65$^{c}$ & 3.62$^{c}$\\
R$_{25}$ (kpc) & 5.3$^{c}$ & 5.7$^{c}$ \\
SFR (M$_{\odot}yr^{ -1}$) & 0.45 & 0.63 \\
log(O/H)+12$_{central}$ & 8.61 & 8.74 \\
%Gradient (r/R$_{25}$) & --0.36$\pm$0.01 & --0.61$\pm$0.22 \\
Gradient (dex kpc$^{-1}$) & --0.07 & --0.11 \\
\vspace{2mm}
log(O/H))+12$_{mean}$ & 8.40 & 8.44 \\
N(O7V) & 4200 & 6000 \\
N(O) & 6250 & 9000 \\
WR candidates & 74 & 160 \\
Spectra obtained & 53\% & 42\% \\
N(WN)$_{\rm spect}$ & 27 & 18 \\
N(WC)$_{\rm spect}$ & 25 & 24 \\
N(WR)$_{\rm total}$ & 105 &  170 \\
N(WR)/N(O) & 0.017 & 0.019 \\
\hline
\multicolumn{3}{l}{$^{a}$\citet{Karachentsev2003}, $^{b}$\citet{Karachentsev2007},}\\
$^{c}$\citet{deVau1991}\\
\end{tabular}
\label{gal_prop}
\end{table}

\subsection{Comparison with evolutionary models}

Evolutionary models of massive stellar populations
\citep{Eldridge2006, Meynet2005} predict a correlation in the WR/O star
and WC/WN star ratios as a function of metallicity, which has been
confirmed by observations of Local Group galaxies \citep{Massey1996}.

Figure \ref{wr_o} compares the empirical N(WR)/N(O) ratio of different
galaxies at a range of metallicities with evolutionary predictions
from \citet{Eldridge2006} and \citet{Meynet2005}. The single and
binary models of \citet{Eldridge2006} which include metallicity
dependent WR winds are a somewhat better overall match to the observed
data than the single star predictions from \citet{Meynet2005}.
However, neither model is successful in tracing the N(WR)/N(O)
  ratio with both appearing to overestimate the number of WR stars,
  for example N(WR)/N(O)\,=\,0.019$^{+0.005}_{-0.003}$ for
NGC 5068 (similar to NGC 7793) is lower than both sets of
predictions. However, as discussed above, if we have underestimated
the early-type WN population, which contribute $\sim$50\% of the WR
stars in the LMC, then the ratio could readily be as high as
N(WR)/N(O)$\sim$0.03 which would be in good agreement with theoretical
predictions of \citet{Eldridge2006}. 

\begin{figure}
\centering
\includegraphics[width=1\columnwidth]{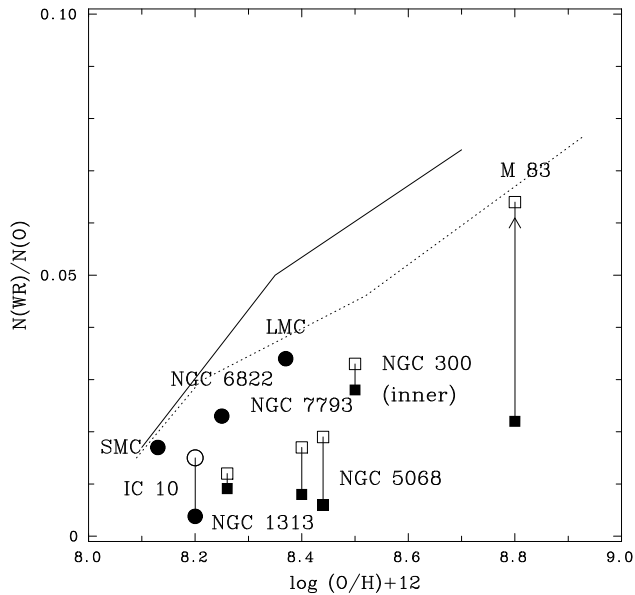}
\caption{Comparison between observed N(WR)/N(O) ratios in nearby and Local
  Group galaxies with predictions from the evolutionary models of
  \citet{Eldridge2006} (dotted) and \citet{Meynet2005} (solid). Filled
  symbols are results from spectroscopy and open symbols account for
  completeness.}
\label{wr_o}
\end{figure}

\section{The Spatial Distribution of WR Stars}

Different core-collapse supernova (ccSN) subtypes appear to be located
in different regions of their host galaxies. For example, Type II SNe,
whose progenitors have been identified as red supergiants (RSG) (See
\cite{Smartt2009} for review), are found to follow the distribution of
the host galaxy light whereas Type Ib, and more so Ic SNe, are
preferentially located in the brightest regions
\citep{Fruchter,Kelly}. Indeed, there is apparently no association
between Type II SNe and H\,{\sc ii} regions according to
\citet{AndersonJames2008}, whereas Type Ib and especially Ic SNe are
intimately associated with bright H\,{\sc ii} regions. In addition,
recent work by \citet{Modjaz2011} has revealed that Type Ic SNe are
located in more metal--rich regions than Type Ib SN. If WN and WC
stars are indeed the progenitors of Type Ib and Ic SNe, respectively,
then they should follow similar distributions.

We use a broad-band image of NGC~5068 from the Digitized Sky Survey
(DSS) since the chip gap in our VLT/FORS1 images would cause the light
distribution of the galaxy to be non-uniform.  We subtract all the
foreground stars before degrading the image, to mimic the spatial
resolution of the SDSS galaxies used in \citet{Kelly}, and remove all
the background light \citep{Leloudas2010}. Although we can assess the
likely WR nature of the remaining WR candidates we only consider the
spatial distribution of spectroscopically confirmed WN and WC stars
that have been detected above the 3$\sigma$ level.

Using an IDL routine we obtain the fractional flux at the location of
each WR star as a function of the number of stars. This is presented
in Figure \ref{fractional_flux}. In addition, we performed a KS-test
to assess the probability that two distributions were from the same
parent population; the results of each comparison are presented in
Table \ref{stats}.

From an initial visual inspection, the distribution of the WN stars
appear to be most consistent with the distribution of the Type II SNe
in the brightest and faintest $\sim$30\% of the galaxy but lie
inbetween the Type Ib and Ic-bl SNe distributions in the moderately
bright regions (30\%--70\% fractional flux). The results
from the KS--test are consistent with the qualitative analysis and
show that the WN stars in NGC~5068 are most likely to be the parent
population of Type Ib and Type II SNe, but does not rule out WN stars
as the progenitors of Type Ic-bl SNe.

One possible explanation for the deviation of the WN from the Type Ib/c
distributions in brighter regions is that, WNE stars with low
$\lambda$4686 excesses are extremely difficult to detect in bright
regions at low spatial resolution as discussed in Section
\ref{completeness}.  A deep, high spatial resolution imaging survey 
may reveal a faint WNE population, in brighter regions, shifting
the distribution and making it more consistent with the Type Ib SN
distribution in the brightest regions.

The KS-test of the WC distribution, like the WN distribution, is most
representative of the Type Ib distribution, which is consistent with a
visual inspection. Again, the distribution deviates in the brightest
20\% of the galaxy, suggesting that incompleteness also plays a,
albeit smaller, role for the WC population. Classically, we would
expect the WC distribution to be most consistent with the Type Ic
distributions, whereas the result presented here indicates that WC
stars are possible progenitors of Type Ib SNe as suggested by
\citep{Georgy2009}.

%However,
%from a visual inspection, the WC distribution deviates from the
%broad-lined Type Ib in the faintest $\sim$45\% of the galaxy where the
%WC stars seem to favour fainter regions. 
%This distribution is more reminiscent of that of Type Ib SNe, which
%would indicate a fainter, less massive, WC population as progenitors
%of Type Ib SNe \citep{Georgy2009}.

Note that both the WN and WC distributions do not favour a single
ccSNe distribution across the entire fractional flux range.  It is
plausible that the WN-Type Ib and WC-Type Ic connections deviate from
a one-to-one correlation, such that some WN and WC stars may
produce Type II and Type Ib SNe, respectively.

However, we admit that the results presented in this section are 
based on small number statistics and to achieve a more significant result 
we require spectroscopic confirmation of the remaining WR candidates
listed in Table \ref{sources}, together with other on-going WR
galaxy surveys.

\begin{figure}
\centering
\includegraphics[width=0.65\columnwidth, angle=-90]{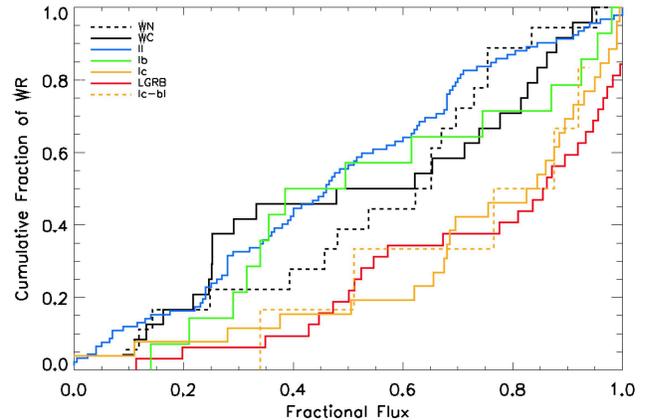}
\caption[Distribution of WR stars in NGC~5068]{A comparison of the
  distribution of WN and WC stars in NGC~5068 (dashed and solid black
  lines respectively) relative to the distribution of light in the
  host galaxy. Plots of supernova and GRB distributions are also
  shown, taken from \citet{Kelly} and \citet{Fruchter},
  respectively.}
\label{fractional_flux}
\end{figure}

\begin{table}
\centering
\caption{Results of the KS test performed on the spatial distribution
  of WR stars in NGC~5068 relative to the galaxy light
  distribution. This reveals that WN stars are most likely the
  progenitors of Type II or Type Ib SNe, while WC stars are more
  likely to produce a broad-lined Ic SNe.}
\begin{tabular}{ccccccc}
\hline
\hline
 & WC & Type II & Type Ib & Type Ic & Type Ic-bl & LGRB \\
\hline
WN & 0.460 & 0.417 & 0.512 & 0.033 & 0.361 & 0.004 \\ 
WC & --      & 0.368  & 0.585 & 0.070 & 0.444 & 0.044 \\
\hline

\end{tabular}
\label{stats}
\end{table}

\section{Summary}

We present results from our VLT/FORS1 imaging and Gemini/GMOS
spectroscopic survey of the WR population in the nearby star-forming
galaxy NGC~5068.

From our VLT/FORS1 narrow-band $\lambda$4684 imaging we have
identified 160 emission line regions, for which we have obtained
photometry. We obtained follow-up Gemini/GMOS spectroscopy 
of 67 of our photometric candidates, of which only 22 exhibited 
statistically significant $\lambda$4684 excesses, while 5 were not
detected in the $\lambda$4781 continuum image. Wolf-Rayet emission
lines were identified in 30 candidates, of which 21 were statistically
significant. Based on line luminosity calibrations based on LMC Wolf-Rayet 
stars we spectroscopically confirmed 18 WN and 24 WC stars.

From comparison of the photometric properties of our confirmed WR
stars with the remaining candidates we infer the presence of at least
an additional 11 WC stars, 43 WN stars and approximately 12 WR stars
of unknown subtype, bringing the total to $\sim$120 Wolf-Rayet stars.
A comparison with the LMC WR stars degraded to a spatial resolution of
$\sim$20 pc suggests that our imaging is fairly complete, although
faint WR sources with intrinsically small m$_{4684}$--m$_{4784}$
excesses, from which we estimate a global WR population of $\sim$170
in NGC~5068.

Our spectroscopy enables us to re-derive the metallicity 
gradient of NGC~5068 using a combination of strong line methods. We find
\[ \log(O/H)+12 =  8.74 \pm 0.15 - (0.61 \pm 0.22) r/R_{25} \]
revealing a strong metallicity gradient, contrary to
 the positive gradient found by \citet{Pilyugin2004}.

 Nebular spectroscopy revealed [N\,{\sc ii}]/H$\alpha$ = 0.155 which
 we applied to the published H$\alpha$ flux of \citet{Kennicutt2008}
 to reveal a
 SFR$\sim$0.63$^{+0.11}_{-0.13}$\,M$_{\odot}$yr$^{-1}$,
 revealing an O star population of
 $\sim$9000$^{+1500}_{-2000}$, and
 N(O)/N(WR)$\sim$0.019$^{+0.005}_{-0.003}$ based on
 N(O)/N(O7V)$\sim$1.5. A comparison between our census of NGC~5068 and
 the survey of NGC~7793 by \citet{bc10} reveals that evolutionary
 predictions may overestimate the N(WR)/N(O) ratio for both galaxies
 by 50\%, unless we have significantly underestimated the fraction of
 faint, weak excess Wolf-Rayet stars.

Finally, we assess the location of our spectroscopically confirmed WN
and WC stars within NGC~5068 relative to the galaxy light and compare
the distributions with those of different subtypes of ccSNe. We find
that both WN and WC stars are most consistent with the Type Ib ccSNe
distribution, however both distributions, particularly WN stars, are
more consistent with the Type II distribution in the brightest
regions. However, for the WN stars this could be explained by the
missing weaker WNE stars which would remain undetected in brighter
regions.

\section*{Acknowledgments}
The authors thank the anonymous referee for careful reading of the
manuscript and for suggestions which helped to improve the quality of
the paper. JLB acknowledges financial support from STFC and also
ongoing support from Hilary Lipsitz. Data presented in this paper are
based on observations made with ESO Telescopes at the Paranal
Observatories under programme ID 081.B-0289 and observations obtained
at the Gemini Observatory, under program ID GS-2009A-Q-20, which is
operated by the Association of Universities for Research in Astronomy,
Inc., under a cooperative agreement with the NSF on behalf of the
Gemini partnership: the National Science Foundation (United States),
the Science and Technology Facilities Council (United Kingdom), the
National Research Council (Canada), CONICYT (Chile), the Australian
Research Council (Australia), Minist\'{e}rio da Ci\^{e}ncia e
Tecnologia (Brazil) and Ministerio de Ciencia, Tecnolog\'{i}a e
Innovaci\'{o}n Productiva (Argentina).

\label{lastpage}

\end{document}